\documentclass[format=acmsmall, review=false, screen=true]{acmart}
\usepackage{booktabs} %
\usepackage{threeparttable}
\usepackage{subfigure}
\usepackage[linesnumbered, ruled]{algorithm2e}

\SetAlFnt{\small}
\SetAlCapFnt{\small}
\SetAlCapNameFnt{\small}
\SetAlCapHSkip{0pt}
\IncMargin{-\parindent}

\acmJournal{TORS}

\usepackage{graphicx}
\usepackage[normalem]{ulem}
\usepackage{xcolor}
\usepackage{amsmath}
\usepackage{multirow}
\usepackage{enumitem}
\usepackage[justification=centering]{caption}

\renewcommand{\thefootnote}{\fnsymbol{footnote}}
\newcommand{\work}[1]{}

\newcommand{\revise}[1]{{#1}}

\begin{document}
	\title{A Survey of Graph Neural Networks for Recommender Systems: Challenges, Methods, and Directions}
	\markboth{~}{Graph Neural Networks for Recommender Systems: Challenges, Methods and Directions}
	\author{Chen Gao*$^{1}$, Yu Zheng*$^{1}$, Nian Li$^{1}$, Yinfeng Li$^{1}$, Yingrong Qin$^{1}$, Jinghua Piao$^{1}$, Yuhan Quan$^{1}$, Jianxin Chang$^{1}$, Depeng Jin$^{1}$, Xiangnan He$^{2}$, Yong Li*$^{1}$}
	\affiliation{%
		\institution{$^{1}$Beijing National Research Center for Information Science and Technology (BNRist),
			Department of Electronic Engineering, Tsinghua University}
		\country{China}
	}
	\affiliation{%
		\institution{$^{2}$School of Information Science and Technology, University of Science and Technology of China}\country{China}
	}
	\email{chgao96@gmail.com, {y-zheng19,linian21,liyf19,qyr16, pojh19,quanyh19,changjx18}@mails.tsinghua.edu.cn, xiangnanhe@gmail.com, {jindp,liyong07}@tsinghua.edu.cn}
	\renewcommand{\shortauthors}{Gao~\textit{et al.}}
	\begin{abstract}
		Recommender system is one of the most important information services on today's Internet. Recently, graph neural networks have become the new state-of-the-art approach to recommender systems. In this survey, we conduct a comprehensive review of the literature on graph neural network-based recommender systems. We first introduce the background and the history of the development of both recommender systems and graph neural networks. For recommender systems, in general, there are four aspects for categorizing existing works: stage, scenario, objective, and application. For graph neural networks, the existing methods consist of two categories, spectral models and spatial ones. We then discuss the motivation of applying graph neural networks into recommender systems, mainly consisting of the high-order connectivity, the structural property of data, and the enhanced supervision signal. We then systematically analyze the challenges in graph construction, embedding propagation/aggregation, model optimization, and computation efficiency. Afterward and primarily, we provide a comprehensive overview of a multitude of existing works of graph neural network-based recommender systems, following the taxonomy above. Finally, we raise discussions on the open problems and promising future directions in this area. We summarize the representative papers along with their code repositories in \url{https://github.com/tsinghua-fib-lab/GNN-Recommender-Systems}.
	\end{abstract}
	\keywords{Recommender Systems, Graph Neural Networks, Graph Representation Learning; Information Retrieval}
	
	\setcopyright{acmlicensed}
	\acmJournal{TORS}
	\acmYear{2023} \acmVolume{1} \acmNumber{1} \acmArticle{1} \acmMonth{1} \acmPrice{15.00}\acmDOI{10.1145/3568022}
	
	\maketitle
	\renewcommand{\thefootnote}{\fnsymbol{footnote}}
	\footnotetext[1]{Yong Li is the \textit{Corresponding Author}. The first two authors contributed equally to this paper. }
	
	\section{Introduction }\label{sec:introduction}
Recommender system, is a kind of filtering system in which the goal is to present personalized information to users, which improves the user experience and promotes business profit.
As one of the typical applications of machine learning driven by the real world, it is an extremely hot topic in both industrial and academia nowadays.

To recap the history of recommender systems, it can be generally divided into three stages, shallow models~\cite{rendle2009bpr,koren2009mf,rendle2010factorization}, neural models~\cite{he2017neural,guo2017deepfm,cheng2016wide}, and GNN-based models~\cite{ying2018graph,wang2019ngcf,he2020lightgcn}.
The earliest recommendation models captured the collaborative filtering (CF) effect by directly calculating the similarity of interactions. Then model-based CF methods, such as matrix factorization (MF)~\cite{koren2009mf} or factorization machine\cite{rendle2010factorization}, were proposed to approach recommendation as a representation learning problem.
However, these methods are faced with critical challenges such as complex user behaviors or data input.
To address it, neural network-based models~\cite{he2017neural,guo2017deepfm,cheng2016wide} are proposed. For example, neural collaborative filtering (NCF) was developed to extend the inner product in MF with multi-layer perceptrons (MLP) to improve its capacity.
Similarly, deep factorization machine (DeepFM)~\cite{guo2017deepfm} combined the shallow model factorization machine (FM)~\cite{rendle2010factorization} with MLP.
However, these methods are still highly limited since their paradigms of prediction and training ignore the high-order structural information in observed data.
For example, the optimization goal of NCF is to predict user-item interaction, and the training samples include observed positive user-item interactions and unobserved negative user-item interactions.
It means that during the parameter updating for a specific user, only the items interacted by him/her are involved.

Recently, the advances in graph neural networks provide a strong and fundamental opportunity to address the above issues in recommender systems.
Specifically, graph neural networks adopt embedding propagation to aggregate neighborhood embedding iteratively. By stacking the propagation layers, each node can access high-order neighbors' information, rather than only the first-order neighbors' as the traditional methods do.
With its advantages in handling the structural data and exploring structural information, GNN-based methods have become the new state-of-the-art approaches in recommender systems.

To well apply graph neural networks into recommender systems, there are some critical challenges required to be addressed.
First, the data input of recommender system should be carefully and properly constructed into graph, with nodes representing elements and edges representing relations.
Second, for the specific task, the components in graph neural networks should be adaptively designed, including how to propagate and aggregate, in which existing works have explored various choices with different advantages and disadvantages.
Third, the optimization of the GNN-based model, including the optimization goal, loss function, data sampling, etc.~\cite{yang2022region}, should be consistent with the task requirement.
Last, since recommender systems have strict limitations on the computation cost, and also due to GNNs' embedding propagation operations introducing a number of computations, the efficient deployment of graph neural networks in recommender systems is another critical challenge.

In this paper, we aim to provide a systematic and comprehensive review of the \revise{research efforts}, especially on how they improve recommendation with graph neural networks and address the corresponding challenges. To fulfill a clear understanding, we categorize research of recommender systems from four perspectives, stage, scenario, objectives, and applications.
We summarize the representative papers along with their code repositories in \url{https://github.com/tsinghua-fib-lab/GNN-Recommender-Systems}.

It is worth mentioning that there is one existing survey~\cite{wu2020graph} of graph neural network-based recommender system, compared with which our differences are as follows.
First, it mainly considers collaborative filtering, sequential recommendation, social recommendation, and knowledge-based recommendation, which belong to different aspects based on the taxonomy and thus cannot be listed side by side.
For example, the sequential recommendation is only one specific recommendation scenario with a special setting of input and output, as pointed out by this survey.
Second, important recommendation stages (such as ranking), recommendation scenarios (such as cross-domain recommendation), and beyond-accuracy recommendation objectives are ignored.
Especially, the beyond-accuracy recommender system is now a very important topic in both academia and industry.
Last, since this area is developing rapidly, there are many recent papers in top-tier venues not covered by~\cite{wu2020graph}.
It is worth mentioning that there is a recently-published book chapter~\cite{chu2022graph} which introduces GNN's basic concepts and several representative applications in modern recommender systems.

Furthermore, in addition to~\cite{wu2020graph}, there are some other surveys closely relevant to this survey as follows.
\begin{itemize}
	\item~\cite{liu2022survey} is a very recent survey on recommender systems based on heterogeneous information network (HIN), which divides literature into three categories: similarity measurement, matrix factorization, and graph representation learning. As for the category of graph representation learning, the authors further divide it into two sub-categories: two-stage training-based methods and end-to-end training-based methods (including relation-based ones and meta-path-based ones). As for the end-to-end methods, this survey briefly introduces some graph neural network-based methods. That is, ~\cite{liu2022survey} only covers a very small fraction of GNN-based recommenders.
	\item~\cite{wang2021graph} is a survey on graph-based recommender systems. It is very short (9 pages) and only uses one page to briefly introduce very-limited typical works of GNN-based recommenders.  ~\cite{deng2021recommender} is a survey of recommender systems based on graph embedding techniques. It mainly discusses the traditional graph embedding methods and only discusses a few GNN-based methods.
	\item~\cite{wu2022survey} is a very recent survey on deep learning-based recommender systems. It mainly discusses two parts of works: deep learning-based collaborative filtering and deep learning-based feature learning for recommendation.
	In the first part, the authors have introduced some representative GNN-based collaborative filtering methods; in the second part, the authors introduced a few works that use GNN to extract visual features. 
	\item  There are other surveys on specific topics of recommendation, such as ~\cite{wang2021survey} for session-based recommendation, ~\cite{guo2020survey} ~\cite{chicaiza2021comprehensive} for knowledge-graph based recommendation, ~\cite{chen2022measuring} for explainable recommendation, ~\cite{wu2022personalized} for news recommendation. There are some other surveys focusing on specific techniques of recommendation, such as automated machine learning~\cite{zheng2022automl} and self-supervised learning~\cite{zheng2022automl}.
\end{itemize}
In summary, our survey is largely different compared with these surveys, as our survey provides extensive and up-to-date introduction and discussions about GNN-based recommenders.

The structure of this survey is organized as follows.
We first introduce the background of recommender systems, from four kinds of perspectives (stage, scenario, objective, application), and the background of graph neural networks, in Section~\ref{sec::background}.
We then discuss the challenges of applying graph neural networks to recommender systems from four aspects, in Section~\ref{sec::challenges}.
Then we elaborate on the representative methods of graph nerual network-based recommendation in Section~\ref{sec::existing-works} by following the taxonomy in the above section. 
We discuss the most critical open problems in this area and provide ideas of the future directions in Section~\ref{sec::open-problems} and conclude this survey in Section~\ref{sec::conclusion}.
	\section{Background}\label{sec::background}

\subsection{Recommender Systems}

\begin{figure*}[t!]
	\centering
	\includegraphics[width=0.8\textwidth]{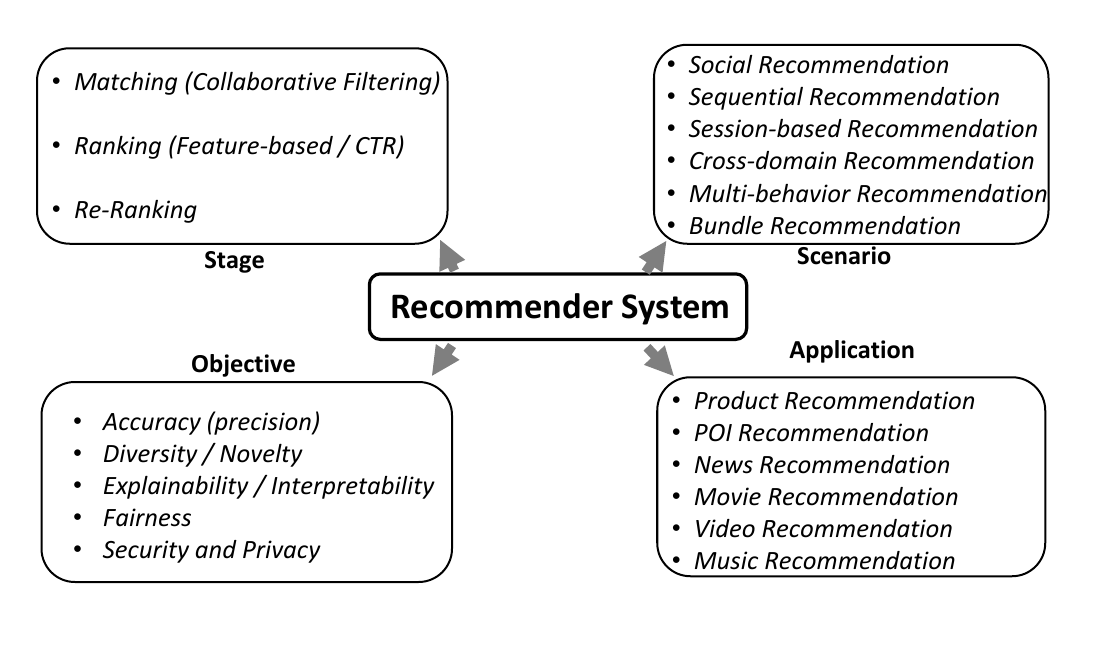} 
	\caption{An illustration of typical recommender systems (stages, scenarios, objectives, and applications)} 
	\vspace{-0.3cm}
	\label{fig:recsys}
\end{figure*}

\subsubsection{Overview}
In this section, we present the background of recommender systems from four perspectives: stages, scenarios, objectives, and applications.
Specifically, in industrial applications, due to the real-world requirements of system engineering, the recommender systems are always split into three stages, \textit{matching}, \textit{ranking}, and \textit{re-ranking}, forming a standard pipeline. Each stage has different characteristics on data input, output, model design, etc.
Besides the standard stages, there are many specific recommendation scenarios with a special definition.
For example, in the last twenty years, social recommendation has been attracting attention, defined as improving recommender systems based on social relations.
Last, different recommender systems have different objectives, of which accuracy is always the most important one as it directly determines the system's utility.
Recently, recommender systems have been assigned other requirements such as recommending diversified items to avoid boring user experience, making sure the system treats all users fairly, protecting user privacy from attack, etc.
As for the applications, GNN models can be widely deployed in e-commerce recommendation, point-of-interest recommendation, news recommendation, movie recommendation, music recommendation, etc.
\subsubsection{Stages }

The item pool,~\textit{i.e.} all the items available for recommender systems, is usually large and can include millions of items. Thus, common recommender systems follow a \textit{multi-stage} architecture, filtering items stage by stage from the large-scale item pool to the final recommendations exposed to users, tens of items~\cite{covington2016deep,wilhelm2018practical}.
Generally, a modern recommender system is composed of the following three stages.

\begin{figure*}[t!]
	\centering
	\includegraphics[width=0.78\linewidth]{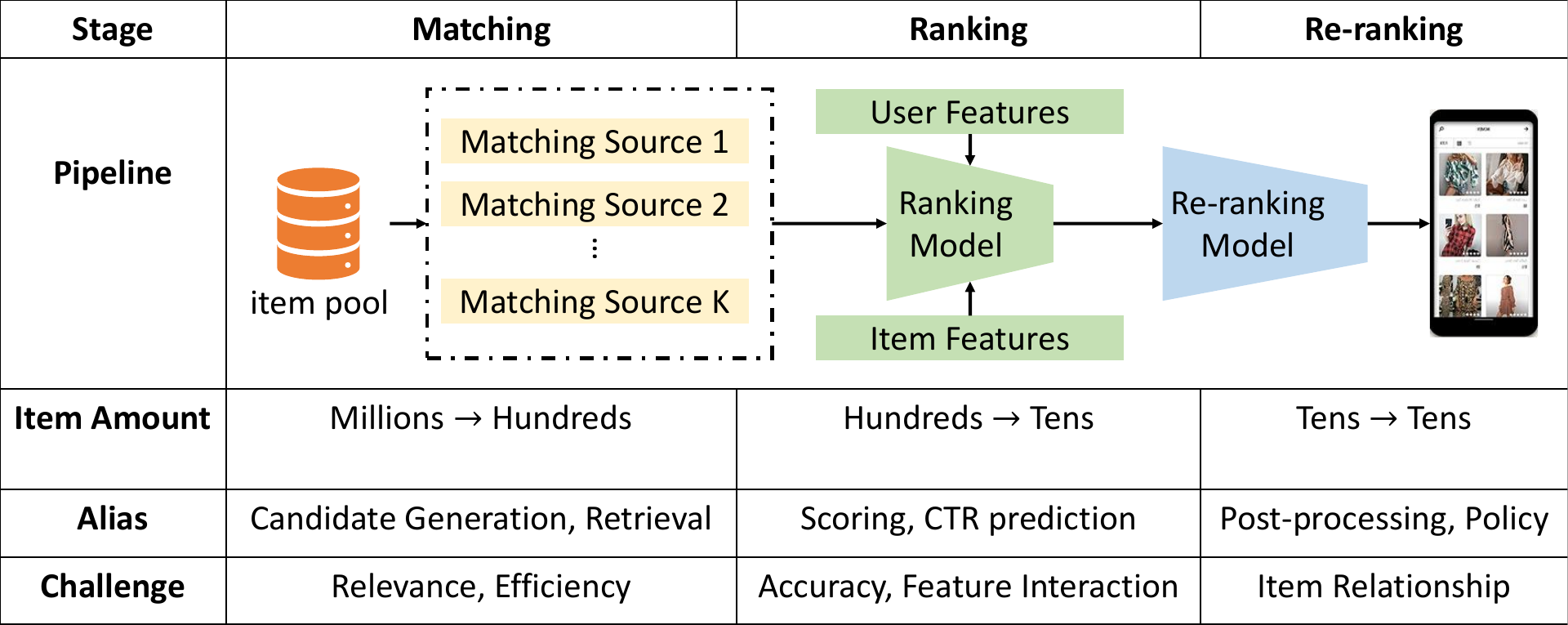} 
	\caption{The typical pipeline of recommender systems.} \label{fig:rs_pipeline}
\end{figure*}

\begin{itemize}[leftmargin=*]
	\item \textbf{Matching.} This first stage generates hundreds of candidate items from the extremely large item pool (million-level or even billion-level), which significantly reduces the scale.
	Considering the large scale of data input in this stage, and due to the strict latency restrictions of online serving, complicated algorithms cannot be adopted, such as very deep neural networks \cite{covington2016deep,kang2019candidate}.
	In other words, models in this stage are usually concise.
	That is, the core task of this stage is to retrieve potentially relevant items with high efficiency and attain coarse-grained modeling of user interests.
	It is worth noting that a recommender system in the real world usually contains multiple matching channels with multiple models, such as embedding matching, geographical matching, popularity matching, social matching, etc.
	
	\item \textbf{Ranking.} After the matching stage, multiple sources of candidate items from different channels are merged into one list and then scored by a single ranking model.
	Specifically, the ranking model ranks these items according to the scores,
	and the top dozens of items are selected.
	Since the amount of input items in this stage is relatively small, the system can afford much more complicated algorithms to achieve higher recommendation accuracy \cite{kang2018self,song2019autoint,lian2018xdeepfm}.
	For example, rich features including user profiles and item attributes can be taken into consideration, and advanced techniques such as self-attention~\cite{kang2018self} can be utilized.
	Since many features are involved, the key challenge in this stage is to design appropriate models for capturing complicated feature interactions.
	\item \textbf{Re-ranking.} Although the obtained item list after the ranking stage is 
	optimized with respect to relevance, 
	it may not meet other important requirements, such as freshness, diversity, fairness and so on~\cite{pei2019personalized}.
	Therefore, a re-ranking stage is necessary, which usually removes certain items or changes the order of the list to fulfill additional criteria and satisfy business needs.
	The main concern in this stage is to consider multiple relationships among the top-scored items \cite{ai2018learning,zhuang2018globally}.
	For example, similar or substitutable items can lead to information redundancy when they are displayed closely in the recommendations.
\end{itemize}
Fig. \ref{fig:rs_pipeline} illustrates the typical pipeline of a recommender system, as well as the comparisons of the three stages above.

\subsubsection{Scenarios}
In the following, we will elaborate on the different scenarios of recommender systems, including social recommendation, sequential recommendation, session recommendation, bundle recommendation, cross-domain recommendation, and multi-behavior recommendation.
\begin{itemize}[leftmargin=*]
	\item \textbf{Social Recommendation.} %
	\begin{figure}[t!]
		\centering
		\includegraphics[width=0.50\textwidth]{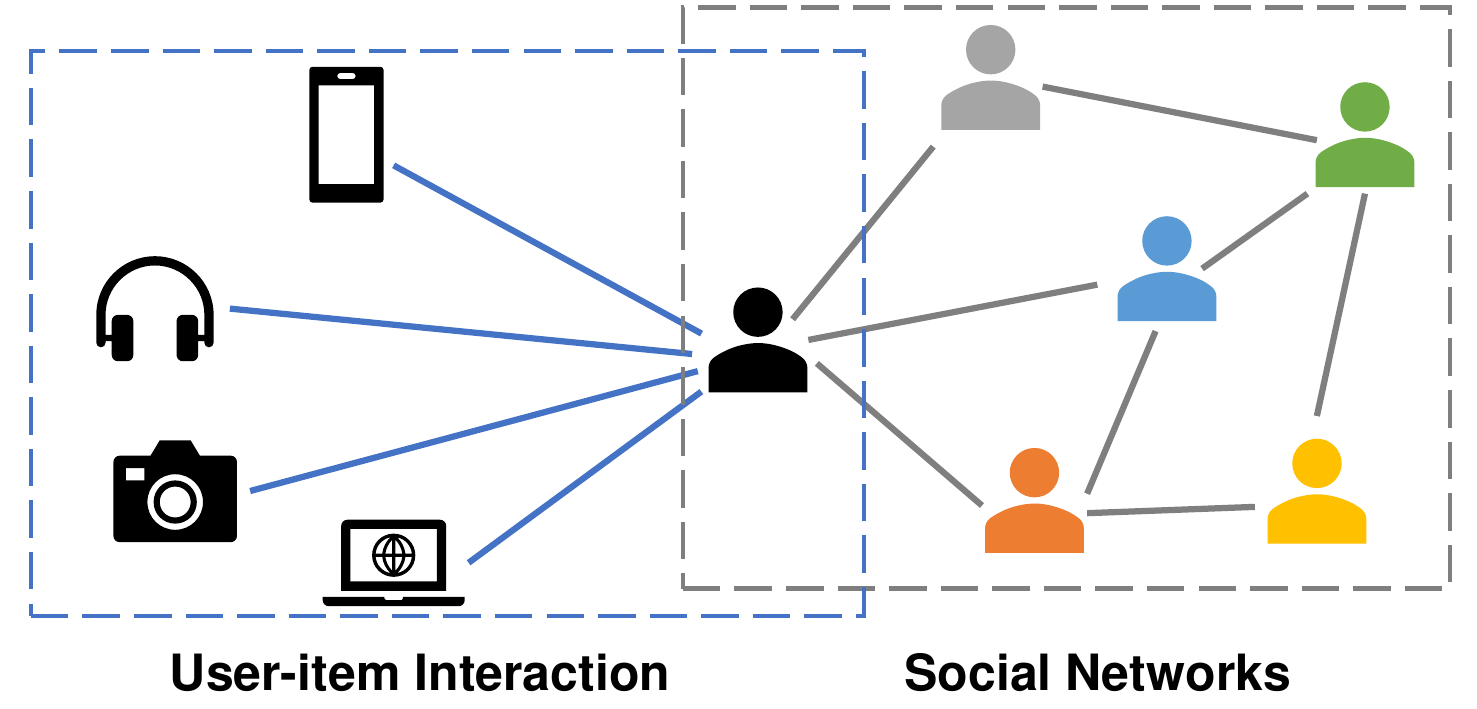} 
		\caption{An illustration of social recommendation. User interactions are affected by both their own preferences and the social factor.} 
		\label{fig:social} 
	\end{figure}
	In the past few years, social platforms have dramatically changed users' daily lives. With the ability to interact with other users, individual behaviors are driven by both personal and social factors.
	Specifically, users' behavior may be influenced by what their friends might do or think, which is known as \textit{social influence}~\cite{cialdini2004social}. For example, users in WeChat\footnote{http://wechat.com/} Video platform may \textit{like} some videos only because of their WeChat friends' \textit{like} behaviors. At the same time, \textit{social homophily} is another popular phenomenon on many social platforms, \textit{i.e.}, people tend to build social relations with others who have similar preferences with them~\cite{mcpherson2001birds}. Taking social e-commerce as an example, users from a common family possibly share similar product preferences, such as food, clothes, daily necessities, and so on.
	Hence, social relations are often integrated into recommender systems to enhance the final performance, which is called social recommendation. Fig.~\ref{fig:social} illustrates the data input of social recommendation, of which user interactions are determined by both preferences and social factors (social influence and social homophily).
	
	\item \textbf{Sequential Recommendation.} %
	\begin{figure}[t]
		\centering
		\includegraphics[width=0.55\textwidth]{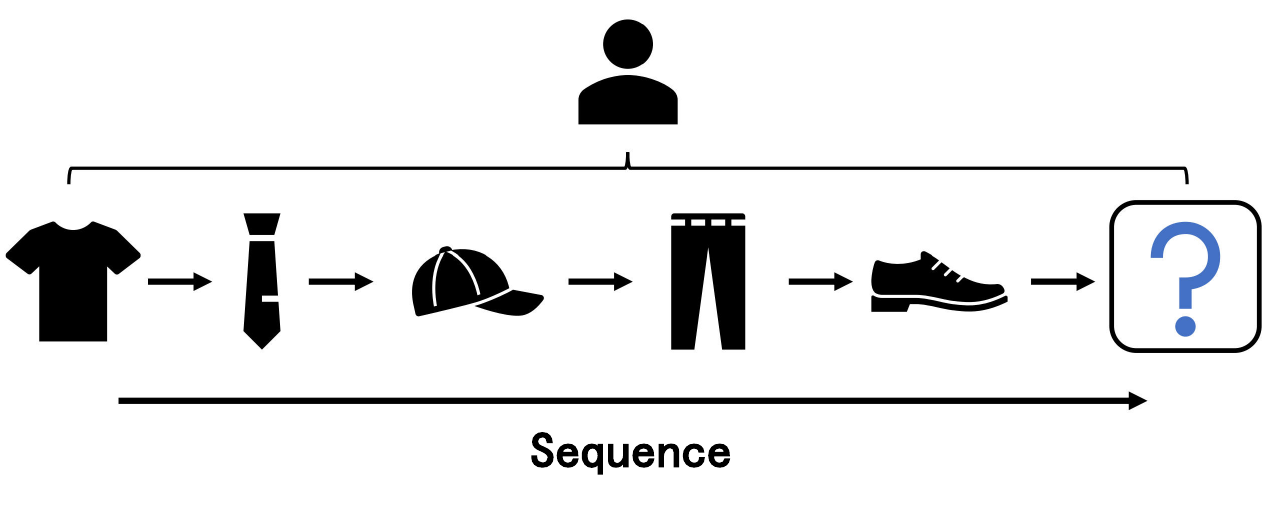} 
		\caption{An illustration of sequential recommendation. Given a user's historical sequence, recommender system aims to predict the next item.}
		\label{fig:sequential}
	\end{figure}
	In recommender systems, users will produce a large number of interaction behaviors over time. The sequential recommendation method extracts information from these behavioral sequences and predicts the user's next interaction item, as shown in Fig.~\ref{fig:sequential}. To symbolize this problem, for the sequence of items $\{x_1, x_2,...,x_n\}$ that the user has interacted with in order, the goal of the system is to predict the next item $x_{n+1}$ that the user will interact with. 
	In recommender systems, the user's historical behaviors play an important role in modeling the user's interest. 
	Many commonly used recommendation methods like collaborative filtering~\cite{he2017neural} 
	train the model by taking each user behavior as a sample. They directly model the user's preference on a single item, but sequential recommendation uses the user's historical behavior sequence to learn timestamp-aware sequential patterns to recommend the next item that the user may be interested in.
	In sequential recommendation, there are two main challenges. First of all, for each sample, \textit{i.e.}, each sequence, the user's interest needs to be extracted from the sequence to predict the next item. Especially when the sequence length increases, it is very challenging to simultaneously model the short-term, long-term, and dynamic interests of users. Secondly, in addition to modeling within a sequence, since items may occur in multiple sequences or users have multiple sequences, collaborative signals between different sequences require to be captured for better representation learning.
	
	\item \textbf{Session-based Recommendation.} %
	\begin{figure}[t]
		\centering
		\includegraphics[width=0.60\textwidth]{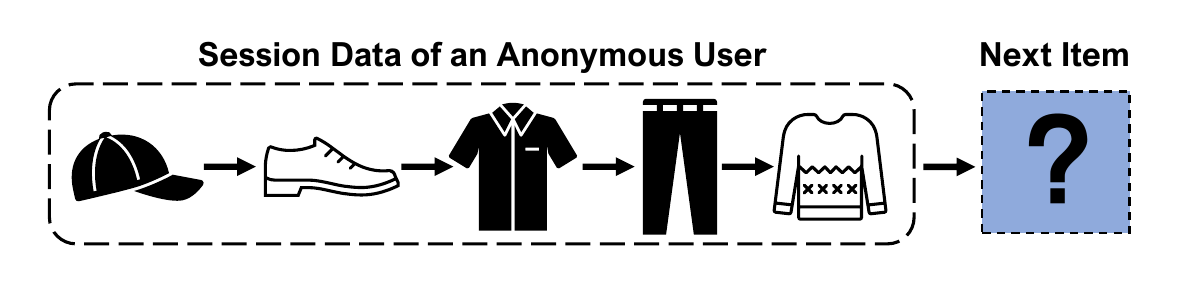} 
		\caption{An illustration of session-based recommendation. Given a anonymous short session, recommender system aims to predict the next item.} \label{fig:session}
	\end{figure}
	In many real-world scenarios, such as some small retailers and mobile stream media (\textit{e.g.} YouTube and Tiktok), it is impossible or not necessary to track the user-id’s behaviors over a long period due to limited storage resources.
	In other words, user profiles and long-term historical interactions are unavailable, and only the short session data from anonymous users are provided. Hence, conventional recommendation methods (\textit{e.g.}, collaborative filtering) may perform poorly in this scenarios. This motivates the problem of session-based recommendation (SBR)~\cite{choi2022s}, which aims at predicting the next item with a given anonymous behavioral session data, as shown in Fig.~\ref{fig:session}.
	Distinct from the sequential recommendation, the subsequent sessions of the same user are handled independently in SBR, since the behavior
	of users in each session only shows session-based traits~\cite{hidasi2015GRU4Rec}.
	\item \textbf{Bundle Recommendation.} %
	\begin{figure}[t]
		\centering
		\includegraphics[width=0.45\textwidth]{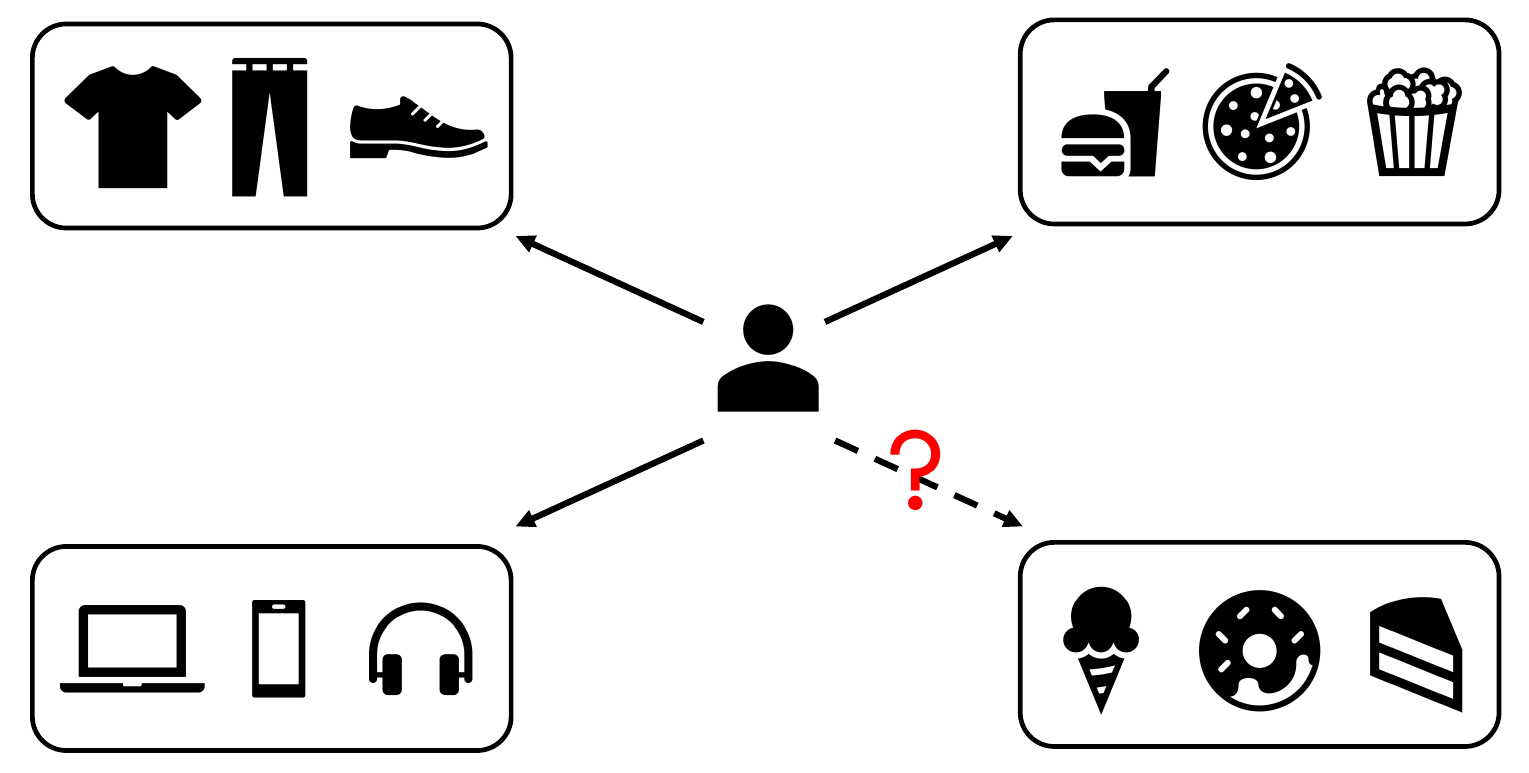} 
		\caption{An illustration of bundle recommendation. } 
		\label{fig:bundle_rec}
	\end{figure}
	The existing recommender systems mainly focus on recommending independent items to users. The bundle is a collection of items, which is an important marketing strategy for product promotion. Bundle recommendation aims to recommend a combination of items for users to consume~\cite{LIRE, EFM, BBPR, DAM}.
	Bundle recommendation is very common nowadays on online platforms, e.g., the music playlists on Spotify, the pinboards on Pinterest, the computer sets on Amazon, and the furniture suites on IKEA. It is worth mentioning that bundle recommendations are also used to solve interesting and meaningful problems such as fashion outfits~\cite{HFGN} and drug packages~\cite{DPR}.
	\item \textbf{Cross-Domain Recommendation.} %
	\begin{figure}[t]
		\centering
		\includegraphics[width=0.48\textwidth]{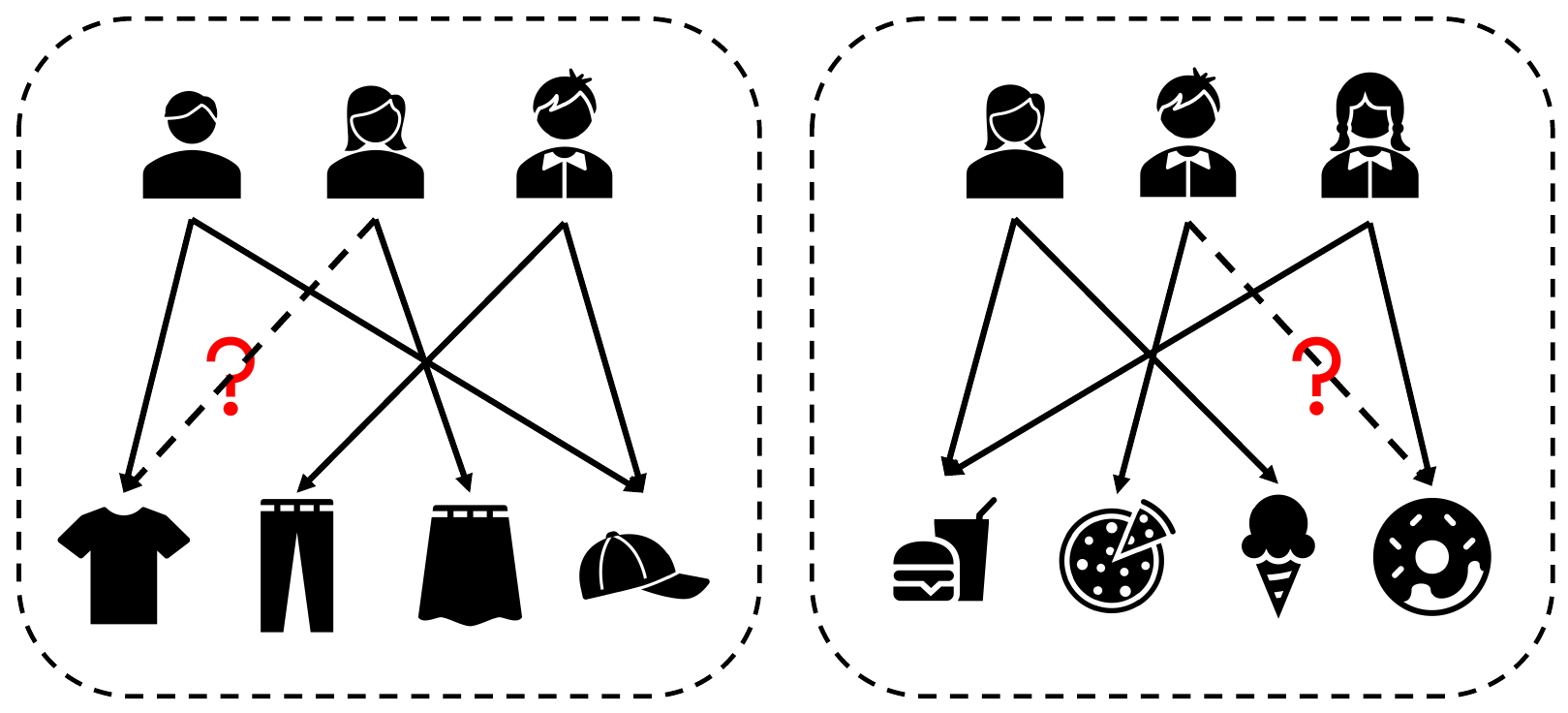} 
		\caption{An illustration of cross-domain recommendation. } 
		\label{fig:cross_domian_rec}
	\end{figure}
	As an increasing number of users interact with multi-modal information across multiple domains, cross-domain recommendation (CDR) has been demonstrated to be a promising method to alleviate cold start and data sparsity problems~\cite{fu2019deeply, gao2019cross, hu2018conet, kang2019semi, man2017cross, mirbakhsh2015improving, zhao2019cross}. 
	CDR methods can be roughly divided into two categories, single-target CDR (STCDR) and dual-target CDR (DTCDR)~\cite{HeroGRAPH}. 
	CDR  methods transfer the information from the source domain to the target domain in one direction; DTCDR emphasizes the mutual utilization of information from both the source domain and target domain, which can be extended to multiple-target CDR (MTCDR). 
	Since utilizing information from multiple domains can improve performance, cross-domain recommendation has become an important scenario in recommender systems.
	
	\item \textbf{Multi-behavior Recommendation.} %
	\begin{figure}[t]
		\centering
		\includegraphics[width=0.50\textwidth]{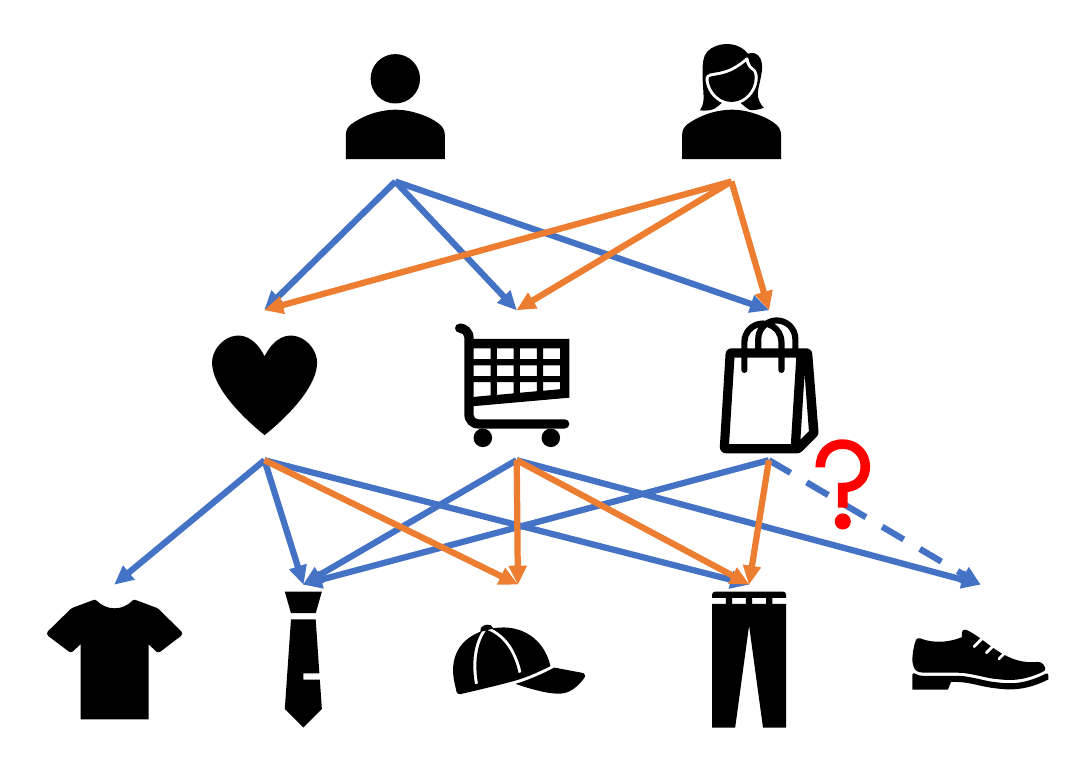} 
		\caption{An illustration of multi-behavior recommendation. } 
		\label{fig:multi_behavior}
	\end{figure}
	Users interact with recommender systems under multiple types of behaviors instead of only one type of behavior. For example, when a user clicks on the video, he/she may also perform behaviors such as collecting or commenting. In an e-commerce website, users often click, add into shopping carts, share, or collect a product before purchasing it, as shown in Fig.~\ref{fig:multi_behavior}. Although the ultimate goal of the recommender system is to recommend products that users will purchase, the purchase behavior is very sparse compared to the user's click, sharing, and other behaviors. 
	To symbolize this problem, For each user $u$ and item $v$, suppose there are $K$ different types of behaviors $\{y_1, y_2,...,y_K\}$. For the $i$-th behavior, if the user has an observed behavior, then $y_i=1$, otherwise $y_i=0$. The goal of the recommender system is to improve the prediction accuracy of a certain type of target behavior $y_t$. 
	
	For multi-behavior recommendation, generally speaking, there are two main challenges. 
	Firstly, different behaviors have different influences on the target behavior. Some behaviors may be strong signals, and some may be weak signals~\cite{xia2021graph, xia2021knowledge}. At the same time, this influence is different for each user. 
	It is challenging to model the influence of these different behaviors on the target behavior accurately.
	Secondly, it is challenging to learn comprehensive representations from different types of behaviors for items. Different behaviors reflect users' different preferences for items; in other words, different behaviors have different meanings. In order to obtain a better representation, the meaning of different behaviors needs to be integrated into the representation learning.

\end{itemize}

\subsubsection{Objectives}
The most important objective of recommender systems, of course, is accuracy.
In the following, we elaborate on the three other important beyond-accuracy objectives, including diversity, explainability, and fairness.

\begin{itemize}[leftmargin=*]
	\item \textbf{Diversity.} %
	\begin{figure}[t]
		\centering
		\includegraphics[width=\linewidth]{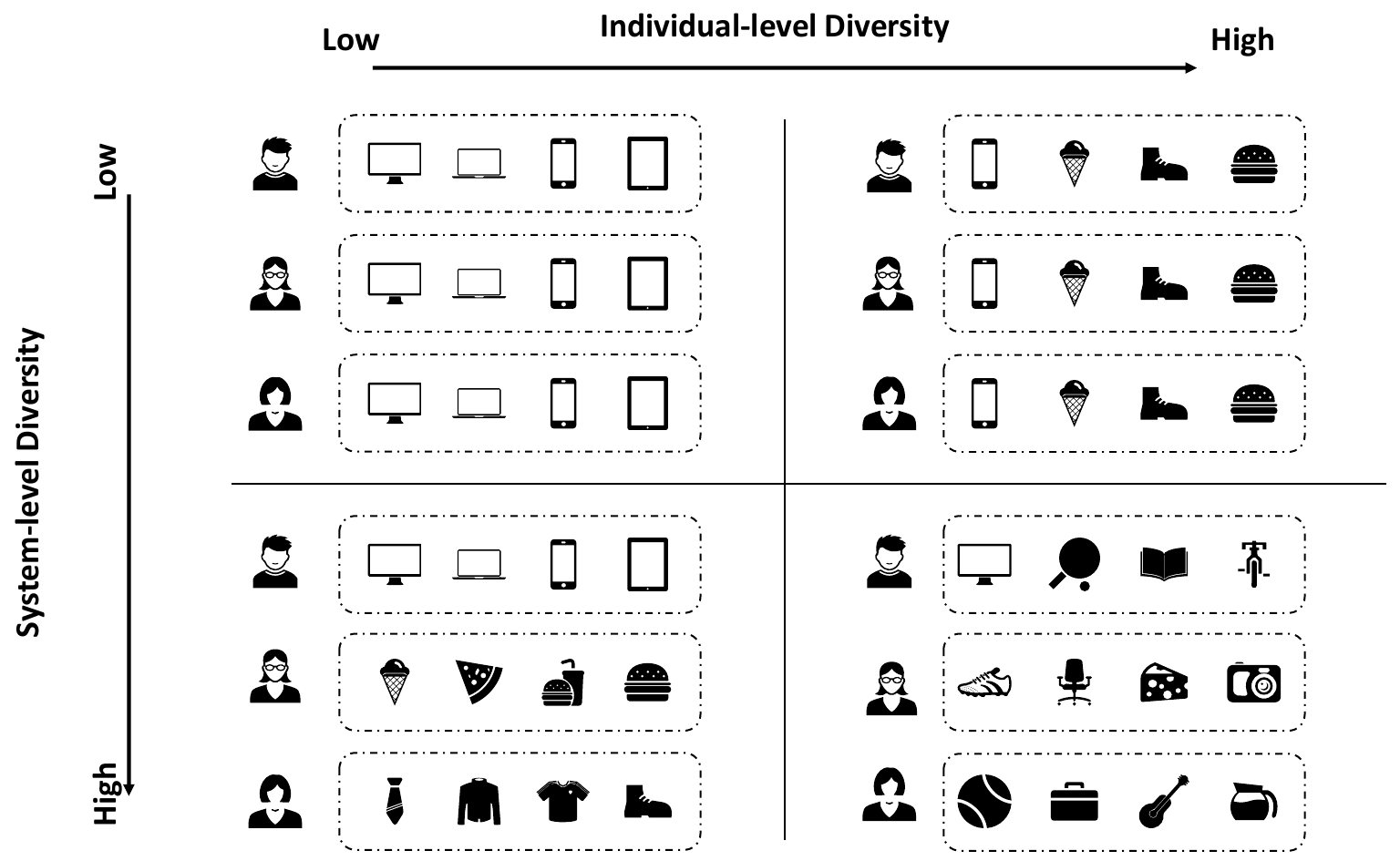} 
		\caption{Illustration of individual-level diversity and system-level diversity.} \label{fig:diversity}
	\end{figure}
	Two types of diversity are usually considered in recommender systems, namely individual-level diversity and system-level diversity.
	Specifically, individual-level diversity is an important objective that measures the dissimilarity of the recommended items for each user since repeated similar items make users reluctant to explore the system.
	In other words, individual-level diversity reflects how many topics the recommendation list covers and how balanced is the recommended items distributed on different topics.
	Here topics depend on the recommendation task, \textit{e.g.} topics can be different product categories for e-commerce recommendation and different genres for music recommendation \cite{zheng2021dgcn}.
	Regarding system-level diversity, it compares the recommendation results of different users and expects them to be dis-similar to each other.
	In other words, low system-level diversity means always recommending popular items to all the users while ignoring long-tail items.
	Therefore, system-level diversity is sometimes called \textit{long-tail} recommendation.
	Fig. \ref{fig:diversity} briefly illustrates the two types of diversity and their differences.
	For both individual-level and system-level diversity, there are two main challenges.
	First, the signal strength of different items varies greatly.
	For each user, there exist dominant topics and disadvantaged topics, \textit{e.g.} a user's interaction records with electronics might be much more frequent than with clothes.
	Similarly, the signal strength of long-tail items is also far weaker than popular items.
	Therefore, it is challenging to recommend relevant content with such weak supervision from either disadvantaged topics or long-tail items for individual-level and system-level diversity, respectively.
	Second, diversity may sometimes contradict recommendation accuracy, resulting in the accuracy-diversity dilemma; thus, it is challenging to balance the two aspects.
	\item \textbf{Explainability.} %
	As current recommender systems mostly adopt a deep learning paradigm, they can arouse urgent needs on the explainability of recommendation~\cite{zhang20explainable}. The focus of explainable recommender systems is not only to produce accurate recommendation results but to generate persuasive explanations for how and why the item is recommended to a specific user~\cite{zhang20explainable,lyu2022knowledge,wang2022multi}. Increasing the explainability of recommender systems can enhance users' perceived transparency~\cite{sinha2002role}, persuasiveness~\cite{tintarev2007survey} and trustworthiness~\cite{kunkel2019let}, and facilitate practitioners to debug and refine the system~\cite{zhang20explainable}. This survey mainly focuses on improving explainability with machine learning techniques. Specifically, past research adopts two different approaches~\cite{zhang20explainable}: one makes efforts to design intrinsic explainable models, ensuring the explainability of recommendation results by designing models with transparent logic (rather than merely ``black box''), e.g., Explicit Factor Models~\cite{zhang2014explicit}, Hidden Factor and Topic Model~\cite{mcauley2013hidden}, and TriRank~\cite{he2015trirank}. The others compromise slightly: they design post-hoc separate models to explain the results generated by ``black box'' recommender systems, e.g., Explanation Mining~\cite{peake2018explanation}. Currently, there are two challenges. First, representing explainable information requires graph-structural item attributes, which are difficult to model without the power of GNN. Second, reasoning recommendations depend on external knowledge in the knowledge graph, which also poses challenges to the task.

	\item \textbf{Fairness.} %
	As a typical data-driven system, recommender systems could be biased by data and the algorithm, arousing increasing concerns on the fairness~\cite{mehrotra18towardsafair, Abdollahpouri20theconnection,li2021tutorial,do2022online}. Specifically, according to the involved stakeholders, fairness in recommender systems can be divided into two categories~\cite{mehrotra18towardsafair,Abdollahpouri20theconnection,li2021tutorial}: user fairness, which attempts to ensure no algorithmic bias among specific users or demographic groups~\cite{leonhardt2018user,li2021user,beutel19fairness}, and item fairness, which indicates fair exposures of different items, or no popularity bias among different items ~\cite{mehrotra18towardsafair,li2021tutorial,abdollahpouri2017controlling,abdollahpouri2019unfairness}. Here, we focus on the user fairness, and leave item fairness in the section of diversity for their close connection in terms of interpretations and solutions~\cite{Mansoury2020fairmatch,abdollahpouri2017controlling}. Specifically, researchers adopt two methods to enhance fairness: One is directly debiasing recommendation results in the training process~\cite{beutel19fairness,zhu2018fairness}, while the other endeavors to rank items to alleviate the unfairness in a post-processing method~\cite{singh2018fairness,Geyik2019Fairness}. Indeed, increasing evidence shows that the utilization of graph data, e.g., user-user, could intensify concerns on fairness~\cite{enyan2021say,Rahman2019fairwalk,wu2021learning}. Thus, it is challenging to debias unfairness in recommendation in the context of rich graph data. Furthermore, it is even harder to boost user fairness in recommendation from the perspective of graph.  
\end{itemize}

\subsubsection{Applications}
Recommender systems widely exist in today's information services, with various kinds of applications, of which the representative ones are as follows.

Product recommendation, also known as E-commerce recommendation, is one of the most famous applications of recommender systems. For recommendation models in the e-commerce scenario, the business value is highly concerned. Therefore, it is crucial to handle multiple types of behaviors closely relevant to the platform profit, including adding-to-cart or purchasing. Some works~\cite{ma2018entire} propose to optimize the click-through rate and conversion rate at the same time.
In addition, in e-commerce platform, products may have rich attributes, such as price~\cite{zheng2020price}, category~\cite{zhou2018DIN}, etc., based on which heterogeneous graphs can be constructed~\cite{luo2020alicoco}. 
\revise{Representative benchmark datasets for product recommendation includes Amazon\footnote{http://jmcauley.ucsd.edu/data/amazon}, Tmall\footnote{https://tianchi.aliyun.com/dataset/dataDetail?dataId=649}, etc.}

POI (Point-of-Interest) recommendation, is also a popular application that aims to recommend new locations/point-of-interests for users' next visitation. 
In the point-of-interest recommendation, there are two important factors, spatial factor, and temporal factor. The spatial factor refers to naturally-existed geographical attributes of POIs, \textit{i.e.}, the geographic location. In addition, the users' visitations are also largely limited by their geographical activity areas since the user cannot visit POIs as easily as browsing/purchasing products on e-commerce websites.
Besides, the temporal factor is also of great importance since users' visitation/check-in behaviors always form a sequence. This motivates the problem of next-POI or successive POI recommendation\cite{lim2020stp,xie2016graph,feng2015personalized,wang2022graph}.
Representative benchmark datasets for POI recommendation includes Yelp\footnote{https://www.yelp.com/dataset}, Gowalla\footnote{https://snap.stanford.edu/data/loc-gowalla.html}, etc.

News recommendation helps users find preferred news, which is also another typical application. Different from other recommendation applications, news recommendation requires proper modeling of the texts of news. Therefore, methods of natural language processing can be combined with recommendation models for extracting news features better~\cite{okura2017embedding}. Besides, users are always interested in up-to-date news and may refuse out-of-date ones. Thus, it is also critical but challenging to fast and accurately filter news from the fast-changing candidate pools.
As for news recommendation, MIND dataset~\cite{wu2020mind} is the recently-released representative benchmark dataset.

Movie recommendation is one of the earliest recommender systems. For example, Netflix's movie recommendation competition~\cite{bell2007lessons} motivated many pioneer recommendation researches~\cite{bell2007lessons,koren2009bellkor}. The earlier setting of movie recommendation is to estimate the users' rating scores on movies, from one to five, which is named \textit{explicit feedback}. Recently, the binary \textit{implicit feedback} has become more popular setting~\cite{he2017neural,rendle2009bpr}.

There are also other types of recommendation applications, such as video recommendation~\cite{davidson2010youtube,liu2020graph,cai2022heterogeneous}, music recommendation~\cite{van2013deep,wang2014improving,la2022music}, job recommendation~\cite{paparrizos2011machine}, food recommendation~\cite{min2019food}, etc.

\subsection{Graph Neural Networks }

\begin{figure*}
	\centering
	\includegraphics[scale=0.42]{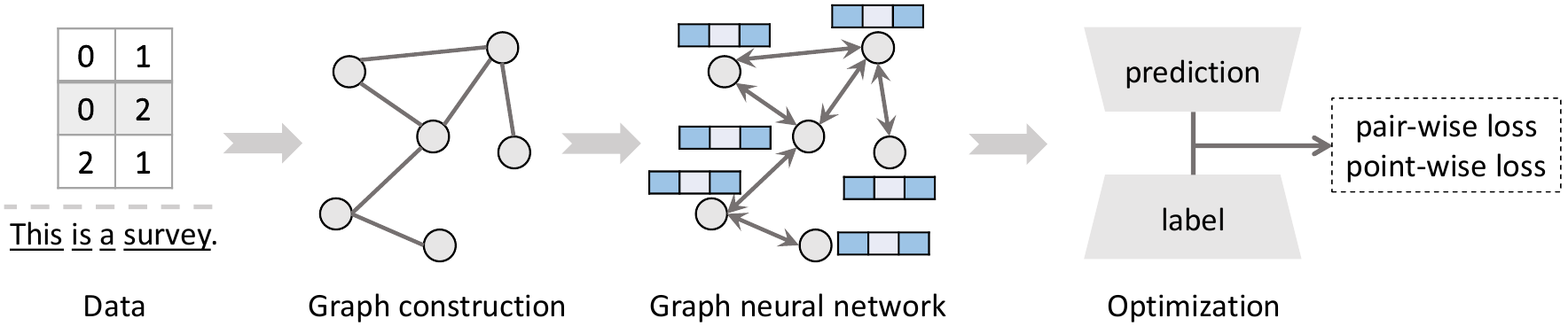}
	\caption{The overall procedure of implementing a GNN model: construct graphs from data~(\textit{e.g.}, table or text), design tailored GNN for generating representations, map representations to prediction results, and further define loss function with labels for optimization.}
	\label{fig:procedure}
\end{figure*}

With the rapid emergence of vast volumes of graph data, such as social networks, molecular structures, and knowledge graphs, a wave of graph neural network (GNN) studies has sprung up in recent years~\cite{kipf2017semi,velivckovic2017gat,berg2018gcmc,zhang2019heterogeneous,fout2017protein,feng2019hypergraph}. The rise of GNN mainly originates from the advancement of convolutional neural network~(CNN) and graph representation learning~(GRL)~\cite{zhou2020graph,wu2020comprehensive}.
When applied to regular Euclidean data such as images or texts, CNN is extremely effective in extracting localized features. However, for non-Euclidean data like graphs, CNN requires generalization to handle the situations where operation objects~(\textit{e.g.}, pixels in images or nodes on graphs) are non-fixed in size. In terms of GRL, it aims to generate low-dimensional vectors for graph nodes, edges, or subgraphs, which represent complex connection structures of graphs. For example, a pioneering work, DeepWalk~\cite{perozzi2014deepwalk}, learns node representations by using SkipGram~\cite{mikolov2013efficient} on a generated path with random walks on graphs. Combining CNN and GRL, various GNNs are developed to distill structural information and learn high-level representations. In later parts, we will introduce several general and primary stages for designing a GNN model to accomplish tasks on graphs, illustrated in Fig. \ref{fig:procedure}. Specifically, section \ref{subsec::construct}, \ref{subsec::design}
and \ref{subsec::opt} elaborate on how to construct graphs, design specialized and effective graph neural networks and optimize models, respectively.

\begin{table*}[t!]
	\caption{Frequently used notations.}
	\begin{tabular}{l|l}
		\hline
		$V$/$E$ &The set of graph nodes/edges  \\
		$\mathcal{U}$/$\mathcal{I}$ &The set of users/items \\
		$\mathcal{N}_i$ &The neighborhood set of graph node $i$ \\
		$\mathbf{h}^l_i$ &The embedding of graph node $i$ in the $l$-th propagation layer \\
		$\mathbf{A}$ & The adjacency matrix of graph \\
		$\mathbf{W}^l$ & Learnable transformation matrix in the $l$-th propagation layer\\
		$\delta(\cdot)$ &Nonlinear activation function \\
		$\Vert$ &Concatenation operation \\
		$\odot$ &Hadamard product \\
		\hline
	\end{tabular}
	\label{tab:notations}
\end{table*}

\subsubsection{Graph Construction}\label{subsec::construct}
We first utilize the unified formula to define a graph, \textit{i.e.}, $\mathcal{G} = (V, E)$, where $V$ and $E$ denote the set of graph nodes and edges respectively, and each edge in $E$ joins any number of nodes in $V$. 
In recent years, GNN-based models focus on designing specialized networks for the following three categories of graphs
\begin{itemize}
	\item \textbf{Homogeneous graph}, of which each edge connects only two nodes, and there is only one type of node and edge.
	\item \textbf{Heterogeneous graph}, of which each edge connects only two nodes, and there are multiple types of nodes or edges.
	\item \textbf{Hypergraph}, of which each edge joins more than two nodes.
\end{itemize}

In many information services nowadays, relational data is naturally represented in the form of graphs. For example, implicit social media relationships can be considered as a unified graph, with nodes representing individuals and edges connecting people who follow each other. However, since non-structured data such as images and texts do not explicitly contain graphs, it is necessary to define nodes and edges manually for building graphs. Taking text data used in Natural Language Processsing~(NLP) as an example, words/documents are described as nodes, and edges among them are constructed according to Term Frequency-Inverse Document Frequency~(IF-ITF)~\cite{yao2019textgraph}. 
An emerging research direction in representation learning on graphs is knowledge graph~(KG), a representative instance of the heterogeneous graph. KG integrates multiple data attributes and relationships, in which nodes and edges are redefined as entities and relations, respectively.
Specifically, the entities in KG can cover a wide range of elements, including persons, movies, books, etc. The relations are utilized to describe how entities associate with each other. For example, a movie can relate to persons~(\textit{e.g.}, actors or directors), countries, languages, \textit{etc.} Except for regular graphs, the hypergraph is also explored recently to handle more complex data~(\textit{e.g.}, beyond pairwise relations and multiple modals) flexibly~\cite{feng2019hypergraph}, in which each edge can connect more than two nodes.

In summary, constructing graphs necessitates either pre-existing graph data or abstracting the concept of graph nodes and edges from non-structured data.

\subsubsection{Network Design}\label{subsec::design}
Generally speaking, GNN models can be categorized into spectral and spatial models. Spectral models consider graphs as signals and process them with graph convolution in the spectral domain. Specifically, graph signals are first transformed into spectral domain by Fourier transform defined on graphs, then a filter is applied, at last the processed signals are transformed back to spatial domain~\cite{shuman2013emerging}. The formulation of processing the graph signal $\mathbf{x}$ with the filter $\mathbf{g}$ is
\begin{equation}
	\mathbf{g} \star \mathbf{x} = \mathcal{F}^{-1}\left(\mathcal{F}(\mathbf{g})\odot \mathcal{F}(\mathbf{x})\right),
\end{equation}
where $\mathcal{F}$ denotes the graph Fourier transform.

In contrast, spatial models conduct the convolution on graph structures directly to extract localized features via weighted aggregation like CNNs. Despite the fact that these two types of models start from different places, they fall into the same principle of collecting neighborhood information iteratively to capture high-order correlations among graph nodes and edges. Here ``information'' is represented as embeddings, \textit{i.e.}, low-dimensional vectors. To this end, the primary and pivotal operation of GNN is to propagate embeddings on graphs following structural connections, including aggregating neighborhood embeddings and fusing them with the target (a node or an edge) embedding to update the graph embeddings layer by layer.

\begin{figure}
	\centering
	\includegraphics[scale=0.45]{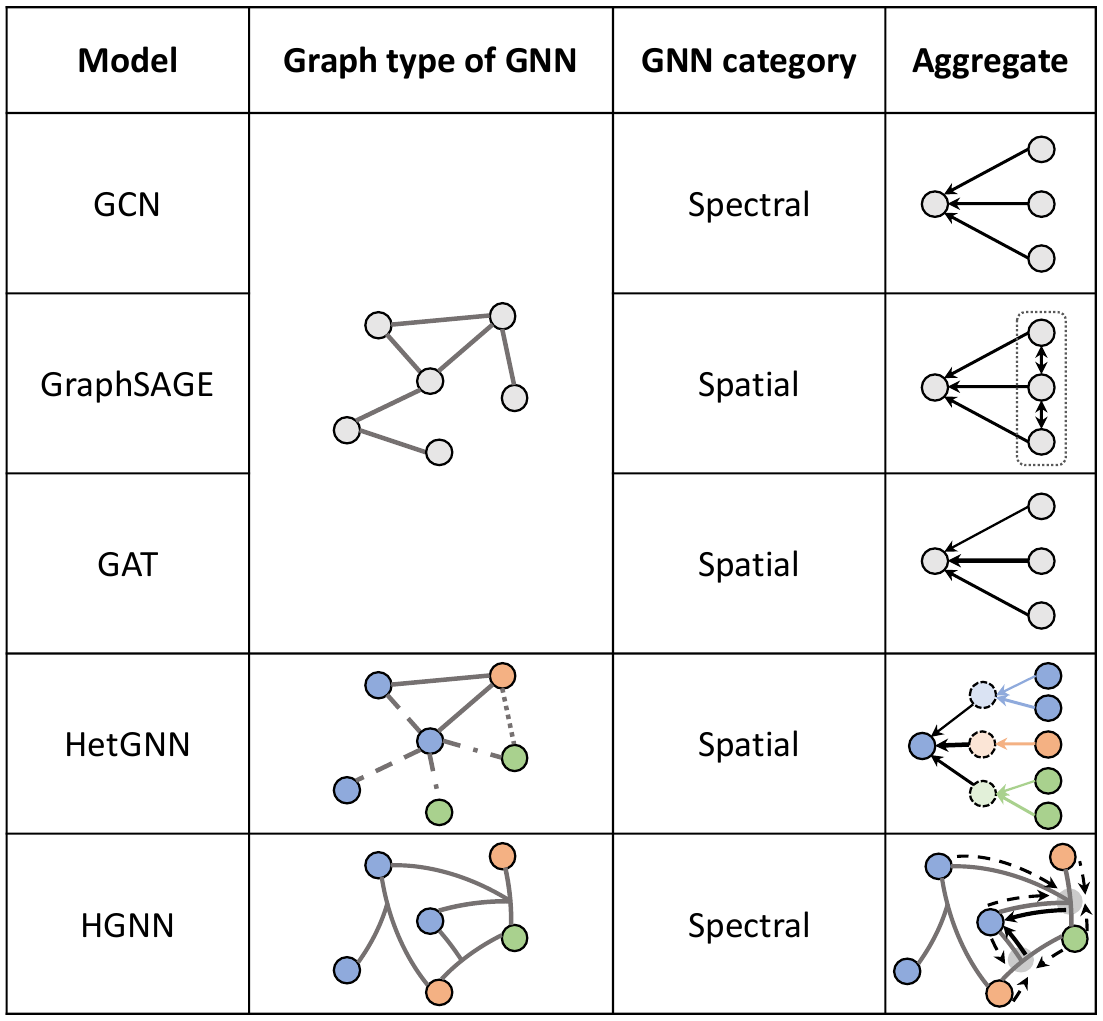}
	\caption{Comparison of several typical GNN models. For graph type, nodes and edges type are represented by colors and line-styles respectively. For aggregation, line-width indicates neighborhood weight.}
	\label{fig:models}
\end{figure}

In the following, we will introduce several groundbreaking GNN models to elaborate how neural networks are implemented on graphs. The frequently used notations are explained in Table \ref{tab:notations}.
\begin{itemize}[leftmargin=*]
	\item \textbf{GCN}~\cite{kipf2017semi}. This is a typical spectral model that combines graph convolution and neural networks to achieve the graph task of semi-supervised classification. In detail, GCN approximates the filter in convolution by the first order following~\cite{hammond2011wavelets}. Then the node embeddings are updated as follows,
	\begin{equation}
		\mathbf{H}^{l+1} = \delta(\tilde{\mathbf{D}}^{-\frac{1}{2}}\tilde{\mathbf{A}}\tilde{\mathbf{D}}^{-\frac{1}{2}}\mathbf{H}^{l}\mathbf{W}^{l}),
	\end{equation}
	of which the derivation can refer to \cite{wu2020comprehensive}.
	$\mathbf{H}^{l}\in \mathbf{R}^{\vert V\vert\times D}$ is the embedding matrix of graph nodes in the $l$-th layer of convolution, where $D$ is the embedding dimension. Besides, $\tilde{\mathbf{A}}\in \mathbf{R}^{\vert V\vert\times\vert V\vert}$ is the adjacency matrix of the graph with self-loop, of which each entry $\tilde{\mathbf{A}}_{ij} = 1$ if the node $i$ connects with $j$ or $i = j$; otherwise $\tilde{\mathbf{A}}_{ij} = 0$, and $\tilde{\mathbf{D}}_{ii} = \sum_j \tilde{\mathbf{A}}_{ij}$.
	
	\item \textbf{GraphSAGE}~\cite{hamilton2017graphsage}. This is a pioneered spatial GNN model that samples neighbors of the target node, aggregates their embeddings, and merges with the target embedding to update.
	
	\begin{equation}
		\begin{aligned}
			&\mathbf{h}_{\mathcal{N}_i}^{l} = \textrm{AGGREGATE}_{l}\left(\{\mathbf{h}_j^{l}, \forall j\in \mathcal{N}_i\}\right), \\
			&\mathbf{h}_i^{l+1} = \delta\left(\mathbf{W}^l\left[\mathbf{h}_i^l\Vert \mathbf{h}_{\mathcal{N}_i}^{l}\right]\right),
		\end{aligned}
	\end{equation}
	where $\mathcal{N}_i$ denotes the sampled neighbors of the target node $i$. The function AGGREGATE has various options, such as MEAN, LSTM~\cite{hochreiter1997lstm} and so on.
	
	\item \textbf{GAT}~\cite{velivckovic2017gat}. This is a spatial GNN model that addresses several key challenges of spectral models, such as poor ability of generalization from a specific graph structure to another sophisticated computation of matrix inverse. GAT utilizes attention mechanisms to aggregate neighborhood features~(embeddings) by specifying different weights to different nodes. Specifically, the propagation is formulated as follows,
	\begin{equation}
		\begin{aligned}
			&\mathbf{h}_i^{l+1} = \delta\left(\sum_{j\in \mathcal{N}_i} \alpha_{ij} \mathbf{W}^l \mathbf{h}_j^{l}\right), \\
			&\alpha_{ij} = \frac{\textrm{exp}\left(\textrm{LeakyReLU}\left(\textbf{a}^T\left[\mathbf{W}^l\mathbf{h}_i^l\Vert \mathbf{W}^l\mathbf{h}_j^l\right]\right)\right)}{\sum_{k\in \mathcal{N}_i}\limits \textrm{exp}\left(\textrm{LeakyReLU}\left(\textbf{a}^T\left[\mathbf{W}^l\mathbf{h}_i^l\Vert \mathbf{W}^l\mathbf{h}_k^l\right]\right)\right)},
		\end{aligned}
	\end{equation}
	where $\alpha_{ij}$ is the propagation weight from node $j$ to node $i$ and $\mathcal{N}_i$ is the neighborhood set of node $i$, including $i$ itself. As shown in the second equation, the attention mechanism is implemented via a fully-connected layer parameterized by a learnable vector $\mathbf{a}$, followed by the \textit{softmax} function. 
	
	\item \textbf{HetGNN}~\cite{zhang2019heterogeneous}. This is a spatial GNN tailored for heterogeneous graphs. Considering the heterogeneous graph consists of multiple types of nodes and edges, HetGNN first divides neighbors into subsets according to their types. Hereafter, an aggregator function is conducted for each type of neighbor to gather localized information, combining LSTM and MEAN operations. Furthermore, different types of neighborhood information are aggregated based on the attention mechanism. Detailed formulas are omitted since the implementation follows the above works.
	
	\item \textbf{HGNN}~\cite{feng2019hypergraph}. This is a spectral model implementing GNN on the hypergraph. The convolution is defined as follows,
	\begin{equation}
		\mathbf{H}^{l+1} = \mathbf{D}_v^{-\frac{1}{2}}\mathbf{E}\mathbf{D}_e^{-1}\mathbf{E}^T\mathbf{D}_v^{-\frac{1}{2}}\mathbf{H}^l\mathbf{W}^l,
	\end{equation}
	where each entry $\mathbf{E}_{ui}$ of $\mathbf{E}\in \mathbf{R}^{\vert V\vert \times \vert E\vert}$ denotes whether the hyperedge $u$ contains the node $i$, each diagonal entry of $\mathbf{D}_v\in \mathbf{R}^{\vert V\vert \times \vert V\vert}$ denotes how many hyperedges the node is included in and each diagonal entry of $\mathbf{D}_e\in \mathbf{R}^{\vert E\vert \times \vert E\vert}$ denotes how many nodes the hyperedge includes. Generally, this convolution operation can be considered as two stages of propagating neighborhood embeddings: \textbf{1)} propagation from nodes to the hyperedge connecting them, and \textbf{2)} propagation from hyperedges to the node they meet.
\end{itemize}
The commonality and difference among typical GNN models above are illustrated in Fig. \ref{fig:models}.

In order to further capture high-order structural information on the graph, the convolution or embedding propagation mentioned above will be performed for $L$ times. In most cases, $L \leq 4$ since GNNs are suffered from the over-smoothing problem that the updated embeddings will be in small fluctuations when the number of propagation layers becomes larger. Relevant studies focusing on developing deep and effective GNN models will be introduced in section \ref{deeper}.

\subsubsection{Model Optimization}\label{subsec::opt}
After the processing of the designed network in section \ref{subsec::design}, overall embeddings of nodes or edges encoding feature semantics as well as graph structures are produced. To perform the downstream graph learning tasks, these embeddings will be further transformed to targets~(\textit{e.g.}, the probability that a node belongs to a class) by general neural networks~(\textit{e.g.}, MLP).

There are mainly classification, prediction and regression tasks on graphs, including three levels: node, edge, and subgraph. Despite the disparity of various tasks, there is a standard procedure for model optimization. Specifically, relevant embeddings will be mapped and come with labels to formulate the loss function, and then existing optimizers are utilized for model learning. Following this process, there are several types of mapping functions~(\textit{e.g.}, MLP, inner product) and loss functions~(\textit{e.g.}, pair-wise, point-wise) to choose for specific tasks. For pair-wise loss function, the discrimination between positive and negative samples is encouraged, and a typical formulation BPR~\cite{rendle2009bpr} is as follows,
\begin{equation}
	\mathcal{L} = \sum_{p, n} -\textrm{ln}\sigma(s(p) - s(n)),
\end{equation} 
where $\sigma(\cdot)$ is the sigmoid function. $p$ and $n$ denote positive and negative samples respectively, and $s(\cdot)$ is for measuring the samples. For point-wise loss function, it includes mean square error~(MSE) loss, cross-entropy loss and so on.

For a better understanding, we take link prediction and node classification tasks as examples to elaborate on how the GNN model is optimized. For link prediction, the likelihood that whether an edge exists between two nodes $i$, $j$ requires definition. Technically, it is usually calculated based on similarity with node embeddings in each layer of propagation:
\begin{equation}
	s\left(i, j\right) = f(\{\mathbf{h}^l_i\}, \{\mathbf{h}^l_j\}),
\end{equation}
where $f(\cdot)$ denotes the mapping function. Furthermore, we can construct the training data as $\mathcal{O} = \{(i, j, k)\}$, consisting of observed positive and randomly-selected negative samples, $(i, j)$ and $(i, k)$, respectively. Specifically, the node $i$ connects with $j$ on the graph but not with $k$. In recommender system, the samples will indicate that user $i$ has interacted with item $j$ but not interacted with item $k$, where $k$ is sampled from all the other items $u$ has not interacted before. Hereafter, if the pair-wise loss BPR is chosen, the optimization object will be
\begin{equation}
	\mathcal{L} = \sum_{(i,j,k)\in \mathcal{O}} -\textrm{ln}\sigma(s(i,j) - s(i,k)).
\end{equation}
In terms of node classification, node embedding will be transformed to a probability distribution representing which class it belongs to, shown as follows,
\begin{equation}
	\mathbf{p}_i = f(\{\mathbf{h}_i^l\}),
\end{equation}
where $\mathbf{p}_i\in \mathbf{R}^{C\times 1}$ is the distribution and $C$ is the number of classes. Similarly, the training data will be $\mathcal{O} = \{(i, \mathbf{y}_i)\}$, where $\mathbf{y}_i\in \mathbf{R}^{C\times 1}$ and $i$ belongs to the class $c$ means $\mathbf{y}_{ic} = 1$; otherwise $\mathbf{y}_{ic} = 0$. In general, the point-wise loss function is usually chosen for the classification task, such as cross-entropy loss which is formulated as
\begin{equation}
	\mathcal{L} = -\sum_{(i, \mathbf{y}_i)\in \mathcal{O}} \mathbf{y}_i^T \log{\mathbf{p}_i}.
\end{equation}

In short, optimization in GNN-based models treats representations generalized by GNNs as input and graph structures~(\textit{e.g.}, edges, node classes) as labels, and loss functions are defined for training.

\subsection{Why are GNNs required for recommender systems } 

Over the past decade, recommender systems have evolved rapidly from traditional factorization approaches to advanced deep neural networks based models.
Particularly, GNN-based recommenders have achieved the state-of-the-art in many aspects, including different recommendaiton stages, scenarios, objectives, and applications.
The success of GNN-based recommenders can be explained from the following three perspectives: (1) structural data; (2) high-order connectivity; (3) supervision signal.

\noindent\textbf{Structural data.} Data collected from online platforms comes in many forms, including user-item interaction (rating, click, purchase, etc.), user profile (gender, age, income, etc.),  item attribute (brand, category, price, etc.), and etc.
Traditional recommender systems are not capable of leveraging those multiple forms of data, and they usually focus on one or a few specific data sources, which leads to sub-optimal performance since much information is ignored.
By expressing all the data as nodes and edges on a graph, GNN provides a unified way to utilize available data.
Meanwhile, GNN shows strong power in learning representations, and thus, high-quality embeddings for users, items, and other features can be obtained, which is critical to the recommendation performance. 

\noindent\textbf{High-order connectivity.} Recommendation accuracy relies on capturing similarity between users and items, and such similarity is supposed to be reflected in the learned embedding space.
Specifically, the learned embedding for a user is similar to embeddings of items that are interacted with by the user.
Furthermore, those items that are interacted with by other users with similar preferences are also relevant to the user, which is known as the collaborative filtering effect, and it is of great importance for recommendation accuracy.
In traditional approaches, the collaborative filtering effect is only implicitly captured since the training data is mainly interaction records that only contain directly connected items.
In other words, only first-order connectivity is taken into consideration. 
The absence of high-order connectivity can largely damage recommendation performance. 
In contrast, GNN-based models can effectively capture high-order connectivity. 
Specifically, the collaborative filtering effect can be naturally expressed as multi-hop neighbors on the graph, and it is incorporated into the learned representations through embedding propagation and aggregation.

\noindent\textbf{Supervision signal.} Supervision signals are usually sparse in the collected data, while GNN based model can leverage semi-supervised signals in the representation learning process to alleviate this problem. 
Take the E-Commerce platform as an example; the target behavior, purchase, is pretty sparse compared to other behaviors. 
Therefore, recommender systems that only use the target behavior may have poor performance.
GNN-based models can effectively incorporate multiple non-target behaviors, such as search and add to cart, by encoding semi-supervised signals over the graph, which can significantly improve recommendation performance \cite{jin2020multi}. 
Meanwhile, self-supervised signals can also be utilized by designing auxiliary tasks on the graph, which further improves recommendation performance.

	\section{Challenges of applying GNNs to recommender systems }\label{sec::challenges}
Although it is well motivated to apply graph neural networks in recommender systems, there exist four parts of critical challenges.

\begin{itemize}
	\item How to construct appropriate graphs for specific tasks?
	\item How to design the mechanism of information propagation and aggregation?
	\item How to optimize the model?
	\item How to ensure the efficiency of model training and inference?
\end{itemize}
In the following, we will elaborate on the four challenges one by one.

\subsection{Graph Construction}

Obviously, the first step of applying graph neural networks is to construct graphs.
This is in two folds: constructing the data input as graph-structured data; reorganizing the recommendation goal as a task on the graph.
Taking the task of standard collaborative filtering as an example,
the data input is the observed user-item interaction data, and the output is predictions of the missing user-item interactions.
Therefore, a bipartite graph with users/items as nodes and interactions as edges can be constructed.
Besides, the CF task turns to the user-item link prediction on the graph.

However, it is challenging to construct graphs that can well handle the task properly.
It should be carefully implemented with the consideration of the following aspects.

\begin{itemize}[leftmargin=*]
	
	\item \textbf{Nodes.}
	One of the main goals of learning with graph neural networks is to assign nodes the representations.
	This results in that the definition of nodes largely determines the scale of the GNN models, of which the majority of parameters are occupied by the layer-$0$ node embeddings. Note that edge embeddings are usually either not considered or computed based on node embeddings. On the other hand, it is also a challenging problem to determine whether to distinguish different types of nodes. For example, in the collaborative filtering task, user and item nodes can be modeled differently or considered the same kind of nodes. Another challenging point is handling concrete input such as numerical features like item prices, which are always continuous numbers.
	To represent these features in the graph, one possible solution is to discretize them into categorical ones, which can be then represented as nodes~\cite{zheng2020price}.

	\item \textbf{Edges.}
	The definition of edges highly affects the graph's quality in further propagation and aggregation, along with the model optimization.
	In some trivial tasks, the recommender system's data input can be considered a kind of relational data, such as user-item interactions or user-user social relations.
	In some complex tasks, other relations can also be represented as edges. For example, in bundle recommendation, a bundle consists of several items. Then the edge connecting the bundle and item can reflect the relation of affiliation. When constructing a graph, good designs of edges should fully consider the graph density.
	A too-dense graph means there are nodes with extremely high degrees. This will make the embedding propagation conducted by a huge number of neighbors.
	It will further make the propagated embedding non-distinguished and useless. 
	To handle too dense edges, sampling, filtering, or pruning on graphs are promising solutions.
	A too-sparse graph will also result in the poor utility of embedding propagation since the propagation will be conducted
	on only a tiny fraction of nodes.

\end{itemize}

\subsection{Network Design}
The propagation layer makes GNN different from the traditional graph learning methods.
As for the propagation, how to choose the path is critical for modeling the high-order similarity in recommender systems.
Besides, the propagation can also be parametric, which assigns different weights to different nodes.
\revise{For example, propagating item embeddings to a user node in the user-item interaction graph captures the item-based CF effect.}
The weights refer to the different importance of historically interacted items.

In the propagation, there are also various choices of aggregation functions, including mean pooling, LSTM, max, min, etc.
Since there is no single choice that can perform the best among all recommendation tasks or different datasets,
it is vital to design a specific and proper one.
Besides, the different choices of propagation/aggregation highly affect the computation efficiency.
For example, mean pooling is widely used in GNN-based recommendation models since it can be computed efficiently, especially for
the graph containing high-degree nodes, such as very popular items (which can connect many users).
Also, the propagation/aggregation layers can be stacked to help nodes access higher-hops neighbors.
Too shallow layers make the high-order graph structure cannot be well modeled, and too deep ones make the node embedding
over-smoothed. Either one of the two cases will lead to poor recommendation performance.

\subsection{Model Optimization}
To optimize the graph neural network-based recommendation models, the traditional loss functions in recommender system
always turn to graph learning losses.
For example, the \textit{logloss} in the optimization can be regarded as the point-wise link prediction loss.
Similarly, BPR loss~\cite{rendle2009bpr} is usually adopted in the link prediction task on graphs.
Another aspect is data sampling. In GNN-based recommendation, to sample positive or negative items, the sampling manner can
highly depend on the graph structure.
For example, in social recommendation, performing random walk on the graph can generate weak positive items (such as items interacted by friends).

In addition, sometimes, GNN-based recommendation may involve multiple tasks, such as the link prediction tasks on different types of edges.
Then in such a case, how to balance each task and make them enhance each other is challenging.

\subsection{Computation Efficiency}

In the real world, recommender systems should be trained/inferred efficiently. Therefore, to ensure the application value
of GNN-based recommendation models, their computation efficiency should be seriously considered.
Compared with
traditional non-GNN recommendation methods such as NCF or FM, GNN models' computation cost is far higher. Complex matrix operations are involved in each GCN layer, especially
for the spectral GNN models such as GCN. With multi-layer stacking of GCN layers, 
the computation cost further increases.
Therefore, spatial GNN models such as PinSage can be easier to be implemented in large-scale industrial applications.
With sampling among neighbors or pruned graph structure, efficiency can always be kept as long as we can bear the drop in recommendation performance.

	\section{Existing Methods}\label{sec::existing-works}

\subsection{Taxonomy~}
In recent years, GNN has been applied to a wide range of recommendation tasks. Here we define the taxonomy in terms of recommendation stages, scenarios, objectives, and applications, respectively. To be more specific, recommendation stages indicate the overall procedure that a recommender system is implemented in the real-world platform. The procedure includes matching for item candidate selection, ranking for capturing user preferences, and re-ranking for other criteria beyond accuracy. The recommendation scenarios include social recommendation, sequential recommendation, cross-domain recommendation, etc. The recommendation objectives incorporate accuracy, diversity, explainability, fairness, and so on, in which accuracy is of the most concern. The recommendation application refers to specific industrial applications. Table \ref{tab::taxonomy_stage}, \ref{tab::taxonomy_scenario}, and \ref{tab::taxonomy_objective} show representative researches of GNN-based recommendation  published in top-tier venues for different recommendation stages, recommendation scenarios, and recommendation objectives, respectively.

\begin{table*}[t!]
	\centering
	\footnotesize
	\caption{A summary of GNN-based models in different recommendation stages in top-tier venues.}
	\label{tab::taxonomy_stage}
	\begin{tabular}{c|c|c|c}
		\hline
		\textbf{Stage} & \textbf{Model} & \textbf{Venue} & \textbf{Year}
		\\
		\hline
		\hline
		 \multirow{10}{*}{Matching} & GCMC \cite{berg2018gcmc} & KDD & 2018 \\
		 & PinSage \cite{ying2018graph} & KDD & 2018 \\
		 & NGCF \cite{wang2019ngcf} & SIGIR & 2019 \\
		 & LightGCN \cite{he2020lightgcn} & SIGIR & 2020 \\
		 & NIA-GCN \cite{sun2020neighbor} & SIGIR & 2020 \\
		 & DGCF \cite{wang2020disentangled} & SIGIR & 2020 \\
		&IMP-GCN~\cite{liu2021interest} & WWW & 2021\\
		& SGL \cite{wu2021self} & SIGIR & 2021 \\
		&HS-GCN~\cite{liu2022hs} & TKDE & 2022\\
		& LGCN~\cite{yu2022low} & AAAI & 2022 \\
		\cline{1-4}
		\multirow{6}{*}{Ranking} & Fi-GNN \cite{li2019fi} & CIKM & 2019 \\
		 & PUP \cite{zheng2020price} & ICDE & 2020 \\
		 & A2-GCN \cite{liu2020a2} & TKDE & 2020 \\
		 & $L_0$-SIGN \cite{su2021detecting} & AAAI & 2021 \\
		 & DG-ENN \cite{guo2021dual} & KDD & 2021 \\
		& TGIN~\cite{jiang2022triangle} & WSDM & 2022 \\
		\cline{1-4}
		 Re-ranking & IRGPR \cite{liu2020personalized} & CIKM & 2020 \\
		\hline
	\end{tabular}
\end{table*}

\begin{table*}[t!]
	\centering
	\footnotesize
	\caption{A summary of GNN-based models in different recommendation scenarios in top-tier venues.}
	\label{tab::taxonomy_scenario}
	\begin{tabular}{c|c|c|c}
		\hline
		 \textbf{Scenario} & \textbf{Model} & \textbf{Venue} & \textbf{Year}\\
		\hline
		\hline 
		\multirow{10}{*}{Social}& DiffNet~\cite{wu2019diffnet} & SIGIR  & 2019 \\
		 & GraphRec~\cite{fan2019graphrec_social} & WWW & 2019\\
		 & DANSER~\cite{wu2019dual_social} & WWW & 2019\\
		 & DGRec~\cite{song2019DGRec}     & WSDM & 2019 \\
		 & HGP~\cite{kim2019HGP}    & RecSys & 2019 \\
		 & DiffNet++~\cite{wu2020diffnet++} & TKDE & 2020 \\
		 & MHCN~\cite{yu2021MHCN_social} & WWW  & 2021 \\
		 & SEPT~\cite{yu2021SEPT_social} & KDD & 2021 \\
		 & GBGCN~\cite{zhang2021GBGCN}    & ICDE & 2021 \\
		 & KCGN~\cite{huang2021KCGN}    & AAAI & 2021 \\
		 & DiffNetLG~\cite{song2021DiffNetLG}      & SIGIR & 2021  \\
		\cline{1-4}
		 \multirow{9}{*}{Sequential} 
		& ISSR \cite{liu2020inter} & AAAI & 2020 \\
		 & MA-GNN \cite{ma2020memory} & AAAI & 2020 \\
		 & STP-UDGAT\cite{lim2020stp} & CIKM & 2020 \\
		 & GPR\cite{chang2020learning} & CIKM & 2020 \\
		 & GES-SASRec\cite{zhu2021graph} & TKDE & 2021 \\
		 & RetaGNN\cite{hsu2021retagnn} & WWW & 2021 \\
		 & TGSRec\cite{fan2021continuous} & CIKM & 2021 \\
	 & SGRec\cite{li2021discovering} & IJCAI & 2021 \\
		   & SURGE \cite{chang2021sequential} & SIGIR & 2021 \\
		\cline{1-4}
		 \multirow{15}{*}{Session} 
		& SR-GNN~\cite{SRGNN}  & AAAI  & 2019 \\
		 & GC-SAN~\cite{xu2019GC-SAN}  & IJCAI  & 2019 \\
		 & TA-GNN~\cite{yu2020tagnn} & SIGIR  & 2020 \\
		 & MGNN-SPred~\cite{wang2020MGNN-SPred} & WWW & 2020 \\
		 & LESSR~\cite{chen2020LESSR} & KDD  & 2020 \\
		 & MKM-SR~\cite{meng2020MKM-SR}  & SIGIR & 2020 \\
		 & GAG~\cite{qiu2020GAG} & SIGIR &2020 \\
		 & GCE-GNN~\cite{GCEGNN} & SIGIR  & 2020 \\
		 & SGNN-HN~\cite{pan2020SGNN-HN} & CIKM & 2020 \\
		 & DHCN~\cite{DHCN} & AAAI & 2021\\
		 & SHARE~\cite{SHARE} & SDM  & 2021\\
		 & SERec~\cite{chen2021SERec} & WSDM  & 2021 \\
		 & COTREC~\cite{xia2021COTREC}  & CIKM & 2021\\
		 & DAT-MID~\cite{chen2021DAT-MDI} & SIGIR & 2021\\
		 & TASRec~\cite{zhou2021TASRec} & SIGIR & 2021\\
		 & G$^{3}$SR~\cite{deng2022g} & TNNLS & 2022\\
		& HG-GNN~\cite{pang2022heterogeneous} & WSDM & 2022\\
		& CGL~\cite{pan2022collaborative} & TOIS  & 2022\\
		\cline{1-4}
		 \multirow{6}{*}{Bundle} & BGCN\cite{BGCN} & SIGIR & 2020 \\
		 & HFGN\cite{HFGN} & SIGIR & 2020 \\
		 & BundleNet\cite{BundleNet} & CIKM & 2020 \\
	 & DPR\cite{DPR} & WWW & 2021 \\
		 & DPG\cite{zheng2022interaction} & TOIS & 2022 \\
		&MIDGN~\cite{zhao2022multi} & AAAI & 2022 \\
		\cline{1-4}
		 \multirow{5}{*}{Cross Domain} & PPGN\cite{zhao2019cross} & CIKM & 2019 \\
		 & BiTGCF\cite{BiTGCF} & CIKM & 2020 \\
		 & DAN\cite{DAN} & CIKM & 2020 \\
		 & HeroGRAPH\cite{HeroGRAPH} & Recsys & 2020 \\
		 & DAGCN\cite{DAGCN} & IJCAI & 2021 \\
		\hline
	\end{tabular}
\end{table*}

\begin{table*}[t!]
	\centering
	\footnotesize
	\caption{A summary of GNN-based models for different recommendation objectives in top-tier venues.}
	\label{tab::taxonomy_objective}
	\begin{tabular}{c|c|c|c}
		\hline
		\textbf{Objective} & \textbf{Model} & \textbf{Venue} & \textbf{Year}
		\\
		\hline
		\hline
	 \multirow{10}{*}{Multi-behavior} & MBGCN \cite{jin2020multi} & SIGIR & 2020\\
		& MGNN-SPred\cite{wang2020MGNN-SPred} & WWW & 2020 \\
		 & MGNN\cite{zhang2020multiplex} & CIKM & 2020 \\
		 & LP-MRGNN\cite{wang2021incorporating} & TKDE & 2021 \\
	 & GNMR\cite{xia2021multi} & ICDE & 2021 \\
		 & MB-GMN\cite{xia2021graph}  & SIGIR & 2021 \\
		 & KHGT\cite{xia2021knowledge} & AAAI & 2021 \\
		 & GHCF\cite{chen2021graph} & AAAI & 2021 \\
		 & DMBGN\cite{xiao2021dmbgn} & KDD & 2021 \\
		\cline{1-4}
		 \multirow{3}{*}{Diversity} & V2HT \cite{li2019long} & CIKM & 2019 \\
		 & BGCF \cite{sun2020framework} & KDD & 2020 \\
		 & DGCN \cite{zheng2021dgcn} & WWW & 2021 \\
		\cline{1-4}
		 \multirow{8}{*}{Explainability} & RippleNet~\cite{wang2018ripplenet}& CIKM & 2018\\
		 & EIUM~\cite{huang2019explainable}& MM & 2019\\
		 & KPRN~\cite{wang2019explainable}& AAAI & 2019\\
		 & RuleRec~\cite{ma2019jointly} & WWW & 2019\\
		 & PGPR~\cite{Xian2019reinforcement}& SIGIR & 2019\\
		 & KGAT~\cite{wang2019kgat} & KDD & 2019\\
		 & TMER~\cite{chen2021temporal} & WSDM & 2021\\
		\cline{1-4}
		 \multirow{2}{*}{Fairness} & FairGo~\cite{wu2021learning} &
		WWW & 2021 \\
		 & FairGNN~\cite{enyan2021say} & WSDM & 2021 \\
		\hline
	\end{tabular}
\end{table*}

\subsection{GNN in Different Recommendation Stages}
\subsubsection{GNN in Matching}

In the matching stage, efficiency is an essential problem because of the high computation complexity for candidate selection. Specifically, only hundreds of items will be selected from the item pool of million magnitudes for the following ranking stage, based on coarse-grained user preferences. Therefore, proposed models in this stage barely leverage user-item interactions as data input for modeling user preferences without introducing additional features such as user ages, item price, browsing time on the application, etc.
\revise{For matching, besides the basic challenges of graph construction and network design, there is also the challenge of CF signal extraction, as shown in Table~\ref{table:matching}.}

GNN-based models in the matching stage can be regarded as embedding matching, usually designing specialized GNN architecture on the user-item bipartite graph~\cite{berg2018gcmc,wang2019ngcf,sun2020neighbor,wang2020disentangled,wu2021self}. Berg \textit{et al.}~\cite{berg2018gcmc} proposed to pass neighborhood messages by summing and assign weight-sharing transformation channels for different relational edges~(\textit{i.e.}, user-item ratings). Wang~\textit{et al.}~\cite{wang2019ngcf} proposed a spatial GNN in recommendation and obtain superior performance compared with conventional CF methods like MF~\cite{koren2009mf} or NCF~\cite{he2017neural}. Sun \textit{et al.}~\cite{sun2020neighbor} argued that simple aggregation mechanisms like sum, mean, or max cannot model relational information among neighbors and proposed a neighbor interaction-aware convolution to address the issue. Wang \textit{et al.}~\cite{wang2020disentangled} developed disentangled GNN to capture independent user intentions, which extends the set of candidate items in matching and guarantees the accuracy simultaneously. Wu \textit{et al.}~\cite{wu2021self} leverage the stability of graph structure to incorporate a contrastive learning framework to assist representation learning. These GNN-based models can capture high-order similarity among users and items as well as structural connectivity. In this way, the semantics that users with similar interactions will have similar preferences are extended through multiple times of information propagation. On the other hand, the training complexity of GNN-based models was demonstrated~\cite{wang2019ngcf,wang2020disentangled} as acceptable and comparable with non-graph models, especially when the transformation matrix is removed~\cite{he2020lightgcn}. Besides, \cite{ying2018graph} showed that the GNN-based model could be applied to web-scale recommender systems in real-world platforms efficiently and effectively, which combines random walk and GraphSAGE~\cite{hamilton2017graphsage} for embedding learning on a large-scale item-item graph. Table \ref{table:matching} shows the commonality and difference among the GNN models in matching
~\cite{liu2021interest} proposed first to filter promising neighbors and then adopt embedding propagation.

\revise{We would like to briefly discuss the advantages and disadvantages of different GNN-based matching methods. These methods with sampling techniques such as~\cite{ying2018graph} can be used for large graphs, and they are more suitable for denser edges, such as item-item relations. Recent advances~\cite{he2020lightgcn} remove non-linear operations to improve the performance further, but on some datasets, the performance of methods with non-linear operations~\cite{wang2019ngcf} are still comparable.}
In a nutshell, GNN can be applied to recommendation tasks effectively, which balances the accuracy and efficiency of generating candidates from the item pool.

\subsubsection{GNN in Ranking}
In the ranking stage, with a much smaller amount of candidate items, more accurate models can be utilized, and more features can be included.
Existing ranking models usually first convert sparse features into one-hot encodings and then transform them into dense embedding vectors.
These feature embeddings are directly concatenated and fed into DNN~\cite{guo2017deepfm,cheng2016wide} or specifically designed models~\cite{rendle2010factorization,he2017neural,song2019autoint} in an unstructured way to estimate the ranking score.
\revise{The main challenge of utilizing GNN for ranking is designing proper structures to capture feature interactions, as shown in Table~\ref{table:ranking}.}
Specifically, GNN-based ranking models usually consist of two components, encoder, and predictor, which address feature interaction from different directions.
On the one hand, special graph structures can be designed to capture the desired feature interactions in the encoder.
On the other hand, feature interaction can be taken into consideration in the predictor, where the ranking score is estimated by integrating different feature embeddings from the GNN encoder.

Li~\textit{et al.}~\cite{li2019fi}~propose Feature Interaction Graph Neural Networks (Fi-GNN), which constructs a weighted fully-connected graph of all the input features.
The encoder in Fi-GNN is composed of a GAT and a GRU, and the predictor is achieved with attention networks.
Zheng \textit{et al.} \cite{zheng2020price,zheng2021incorporating} investigate the influence of price feature in ranking and propose a model called Price-aware User Preference modeling (PUP).
They design an encoder with GCN on a pre-defined heterogeneous graph to capture price awareness, and a two-branch factorization machine is utilized as the predictor.
Liu~\cite{liu2020a2} also approaches the recommendation problem when some item attributes are available and similarly builds item-attributes edges. Then the item embeddings can reflect both user behaviors and attribute information.
Since not all feature interactions are useful, $L_0$-SIGN \cite{su2021detecting} automatically detects beneficial feature interactions and only reserves those edges, resulting in a learned graph which is further fed into a graph classification model to estimate the ranking score.
In addition, Guo \textit{et al.} \cite{guo2021dual} propose DG-ENN with a dual graph of an attribute graph and a collaborative graph, which integrates the information of different fields to refine the embedding for ranking.
Furthermore, SHCF \cite{li2021sequence} and GCM \cite{wu2020gcm} utilize extra nodes and edge attributes to represent item attributes and context information, respectively.
Classical interaction predictors are adopted, such as inner product and FM.
Table \ref{table:ranking} illustrates the differences between the above GNN-based ranking models in terms of the designs of the encoder and predictor.
\revise{Overall, these GNN-based ranking methods can be regarded as a combination of the traditional feature-interaction methods and GNN-based representation learning. As a result, the performance is still constrained by the feature-interaction component. A model that well couples the graph structure, such as high-order paths, into high-order feature extraction is desired.}

\begin{table}[t!]
	\small
	\centering
	\caption{Details of GNN models in matching stage.}\label{table:matching}
	\begin{tabular}{cccc}
		\hline
		Model  & \revise{Graph Construction} & \revise{Network Design} & \revise{Specifial Design for Collaborative Filtering}              \\
		\hline
		GCMC \cite{berg2018gcmc} & user-item & GCN & weight sharing among relations        \\
		NGCF \cite{wang2019ngcf} & user-item & GCN & enhance propagated information \\
		DGCF \cite{wang2020disentangled} & user-item & GAT & disentangled representations \\
		LightGCN \cite{he2020lightgcn} & user-item &  LightGCN & remove transformation and nonlinearity\\
		SGL \cite{wu2021self} & user-item & GCN & self-supervision on graphs \\
		NIA-GCN \cite{sun2020neighbor} & user-item & NIA-GCN & neighbor interaction~(NI) \\
		PinSage \cite{ying2018graph} & item-item & GCN & neighbor sampling\\
		IMP-GCN~\cite{liu2021interest} & user-item & GCN & neighbor sampling \\
		HS-GCN~\cite{liu2022hs} & user-item & GCN & re-define  distance metrics between nodes \\
		LGCN~\cite{yu2022low} & user-item & low-pass GCN & trainable kernels for spectral graph convolution \\
		\hline
	\end{tabular}
\end{table}
\begin{table}[t!]
	\centering
	\small
	\caption{Details of GNN models in ranking stage.}\label{table:ranking}
	\begin{tabular}{cccc}
		\hline
		Model  & \revise{Graph Construction} & \revise{Network Design} & \revise{Feature Interaction}              \\
		\hline
		Fi-GNN \cite{li2019fi} & fully-connected & GAT+GRU & attention        \\
		PUP \cite{zheng2020price} & pre-defined & GCN & two-branch FM  \\
		$L_0$-SIGN \cite{su2021detecting} & learned & SIGN & graph classification \\
		DG-ENN \cite{guo2021dual} & pre-defined &  LightGCN & DNN \\
		SHCF \cite{li2021sequence} & pre-defined & HGAT & inner product \\
		GCM \cite{wu2020gcm} & pre-defined & context GCN & FM \\
		A2-GCN \cite{liu2020a2} & pre-defined & attentive GCN& inner product \\
		TGIN~\cite{jiang2022triangle} &pre-defined &triangle-based GCN &DNN\\
		\hline
	\end{tabular}
\end{table}

\subsubsection{GNN in Re-ranking}
After obtaining the scores of recommended items, top items are further re-ranked with pre-defined rules or functions to improve the recommendation quality.
Specifically, two key factors need to be considered in re-ranking.
First, different items can have mutual influence by certain relationships such as substitutability and complementarity.
Second, different users tend to have distinct preferences, and thus re-ranking can also be personalized.
\revise{GNN provides a unified way to encode both item relationships and user preferences, and thus the major challenge is how to fuse multiple re-ranking goals.}
Liu \textit{et al.} \cite{liu2020personalized} propose a model called IRGPR to accomplish personalized re-ranking with the help of GNN.
They propose a heterogeneous graph to fuse the two information sources, one item relation graph to capture multiple item relationships, and one user-item scoring graph to include the initial ranking scores.
User and item embeddings are obtained after multiple message propagation layers, including global item relation propagation and personalized intent propagation.
The final order of re-ranked items is generated with a feed-forward network. 
\revise{In short, the current works in GNN-based re-ranking only considered a few re-ranking goals, and others remain less explored.}

\subsection{GNN in Different Recommendation Scenarios}

\subsubsection{GNN in Social Recommendation}

In social recommendation, we have social networks that contain the social relations of each user, and the goal is to utilize the local neighbors' preferences for each user in social networks to enhance the user modeling~\cite{wu2018social_collaborative,wu2018neural_social,sun2018attentive_social,chen2019efficient_social}.
From the perspective of representation learning with GNN, there are two key considerations in social recommendation: 1) how to capture the social factor; 2) how to combine the social factor from friends and user preference from his/her interaction behaviors. Here, we first summarize how the existing works capture the social factors from two perspectives, \textit{i.e.}, graph construction and information propagation.
\begin{itemize}[leftmargin=*]
	\item \textbf{Graph construction.} In social-aware recommender systems, a user's final behavior is decided by both the social impacts from friends and his/her preferences. One of the main challenges in social recommendation is constructing a social graph to capture the social influences of friends.
	Generally speaking, a certain user in social networks is not only influenced by his/her friends (the first-order neighbors) but also influenced by friends' friends (the high-order neighbors). 
	\revise{	}
	Given that normal graph can only model the pairwise relations, normal graph-based methods~\cite{fan2019graphrec_social,wu2019diffnet,wu2020diffnet++,wu2019dual_social,yu2021SEPT_social,yu2020ESRF,guo2020GNN-SoR,luo2020ASR,xu2020SR-HGNN,kim2019HGP,mu2019GAT-NSR,song2021DiffNetLG} stacked multiple GNN layers to capture multi-hop high-order social relations. However, stacked GNN layers may suffer from the over-smoothing~\cite{chen2020measuring_oversmoothing} problem, which may lead to significant performance degradation. Hypergraph-based methods, such as MHCN~\cite{yu2021MHCN_social}, propose to model the high-order social relations with hyperedge~\cite{feng2019hypergraph}, which can connect more than two nodes and model the high-order relations in a natural way. HOSR~\cite{liu2020HOSR} recursively propagates embeddings on the social network to reflect the influence of high-order neighbors in the user representations. To further improve the recommendation performance, some works~\cite{xu2019RecoGCN, zhang2021GBGCN, song2019DGRec,huang2021KCGN,bai2020TGRec} introduce side information when constructing the graph. RecoGCN~\cite{xu2019RecoGCN} unifies users, items, and selling agents into a heterogeneous graph to capture the complex relations in social E-commerce. GBGCN~\cite{zhang2021GBGCN} constructs a graph for organizing user behaviors of two views in the group-buying recommendation, of which the \textit{initiator view} contains those initiator-item interactions and the \textit{participant view} contains those participant-item interactions. DGRec~\cite{song2019DGRec} and TGRec~\cite{bai2020TGRec} introduce temporal information of user behaviors into social recommendation. KCGN~\cite{huang2021KCGN} proposes to capture both user-user and item-item relations with the developed knowledge-aware coupled graph.

	\begin{figure*}[t!]
		\centering
		\includegraphics[width=0.8\textwidth]{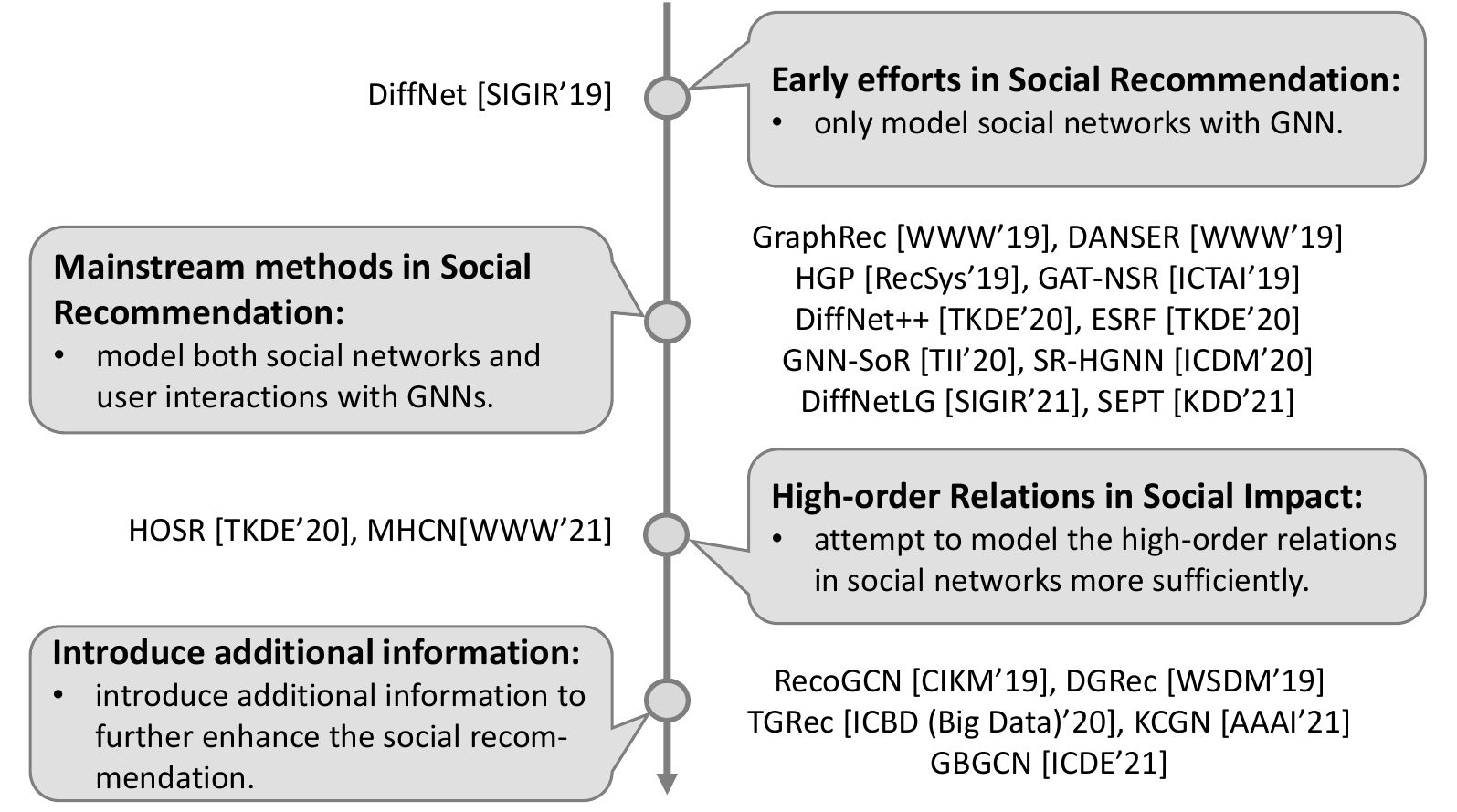} 
		\caption{Illustration of GNN models for social recommendation.} \label{fig:social_map}
	\end{figure*}

	\item \textbf{Information propagation.} 
	As for the propagation on the constructed graph for social recommendation, there are two main propagation mechanisms, \textit{i.e.}, GCN and GAT. Some methods~\cite{wu2019diffnet,yu2021MHCN_social,yu2021SEPT_social,guo2020GNN-SoR,zhang2021GBGCN,xu2020SR-HGNN,huang2021KCGN,kim2019HGP,song2021DiffNetLG,bai2020TGRec} conduct GCN on social graph and treats the social influence of friends equally.
	RecoGCN~\cite{xu2019RecoGCN} conducts meta-path-based GCN propagation on the constructed graph to capture both the social impact and user preference. HOSR~\cite{liu2020HOSR} aggregates the information from neighbors with GCN to capture the high-order relations in the social graph. MHCN~\cite{yu2021MHCN_social} performs propagation with GCN on constructed hypergraph to obtain the high-order social relations.
	Some methods with GAT mechanism~\cite{fan2019graphrec_social,wu2020diffnet++,wu2019dual_social,yu2020ESRF,luo2020ASR,song2019DGRec,mu2019GAT-NSR}, such as GraphRec~\cite{fan2019graphrec_social} and DiffNet++~\cite{wu2020diffnet++}, assume that the social influences from different neighbors on the social graph are different and assign different weights to the social influences from different friends.
\end{itemize}
\begin{table*}[t]
	\centering
	\footnotesize
	\caption{Details of GNN models for social recommendation.}\label{tab:social}
	\begin{tabular}{cccc}
		\hline
		Model & \revise{Graph Construction}  & \revise{Network Design}  & Social Signal Extraction  \\
		\hline
		DiffNet~\cite{wu2019diffnet} & social graph & GCN & sum-pooling \\
		GraphRec~\cite{fan2019graphrec_social} & social graph + user-item graph  & GAT & concatenation\\
		DANSER~\cite{wu2019dual_social} & social graph + user-item graph & GAT \& GCN & - \\
		DiffNet++~\cite{wu2020diffnet++} & heterogeneous graph  & GAT & multi-level attention network \\
		MHCN~\cite{yu2021MHCN_social} &multi-channel hypergraph + user-item graph & HyperGCN & sum-pooling \\
		SEPT~\cite{yu2021SEPT_social} & triangle-graphs + user-item graph & GCN & - \\
		RecoGCN~\cite{xu2019RecoGCN} & heterogeneous graph  & Meta-path + GCN & concatenation \\
		ESRF~\cite{yu2020ESRF}   & motif-induced graph  & GAT  &  sum-pooling   \\
		GNN-SoR~\cite{guo2020GNN-SoR}  & heterogeneous graph & GCN & concatenation   \\
		ASR~\cite{luo2020ASR}   & heterogeneous graph & GAT & concatenation  \\
		GBGCN~\cite{zhang2021GBGCN}   & heterogeneous graph & GCN  & - \\
		DGRec~\cite{song2019DGRec}  & social graph & GAT & -  \\
		SR-HGNN~\cite{xu2020SR-HGNN} & social graph + user-item graph  & GCN & concatenation\\
		KCGN~\cite{huang2021KCGN} & social graph + item-item graph & GCN & concatenation  \\
		HGP~\cite{kim2019HGP} & group-user graph + user-item graph  & GCN & attention mechanism  \\
		GAT-NSR~\cite{mu2019GAT-NSR}  & social graph + user-item graph  & GAT & MLP\\
		HOSR~\cite{liu2020HOSR} & social graph + user-item graph  & GCN  &  attention mechanism   \\
		DiffNetLG~\cite{song2021DiffNetLG}   & heterogeneous graph  & GCN  & concatenation  \\
		TGRec~\cite{bai2020TGRec}  & heterogeneous graph  & GCN  &  attention mechanism \\
		
		\hline
	\end{tabular}
\end{table*}

In social recommendation, user representations are learned from two distinct perspectives, \textit{i.e.} social influence and user interactions. To combine the user representations from the above two perspectives, there are two strategies, 1) separately learn user representations from the social graph and user-item bipartite graph and 2) jointly learn user representations from
the unified graph that consists of social graph and user-item bipartite graph. The methods with the first strategy, such as DiffNet~\cite{wu2019diffnet}, GraphRec~\cite{fan2019graphrec_social} and MHCN~\cite{yu2021MHCN_social}, first separately learn user representations from social graph and user-item graph, and then combines the representations with sum-pooling~\cite{wu2019diffnet,yu2021MHCN_social}, concatenation~\cite{fan2019graphrec_social}, MLP~\cite{mu2019GAT-NSR} or attention mechanism~\cite{kim2019HGP,liu2020HOSR,bai2020TGRec}.
DiffNet++~\cite{wu2020diffnet++}, a typical method with the second strategy, first aggregates the information in the user-item sub-graph and social sub-graph with the GAT mechanism and then combines the representations with the designed multi-level attention network at each layer. Table \ref{tab:social} shows the differences among the above approaches for social recommendation.

To sum up, the development of social recommendation with GNN can be summarized in Fig.~\ref{fig:social_map}. Early efforts in social recommendation only model the social network with GNN, such as DiffNet~\cite{wu2019diffnet}. Then, the methods~\cite{fan2019graphrec_social,wu2019diffnet,wu2020diffnet++,wu2019dual_social,yu2021SEPT_social,yu2020ESRF,guo2020GNN-SoR,luo2020ASR,xu2020SR-HGNN,kim2019HGP,mu2019GAT-NSR,song2021DiffNetLG} that model both social network and user interactions with GNNs become the mainstream in GNN-based social recommendation. Moreover, some studies, such as MHCN~\cite{yu2021MHCN_social} and HOSR~\cite{liu2020HOSR}, attempt to enhance the recommendation by modeling the high-order relations in social networks more sufficiently. Also, there exist some works~\cite{zhang2021GBGCN,song2019DGRec,xu2019RecoGCN,bai2020TGRec,huang2021KCGN} that introduce additional information to further enhance the social recommendation.
\revise{In short, the recent advances that can distinguish social relations with different strengths can always achieve better performance than early efforts roughly treating all social relations as the same.}

\subsubsection{GNN in Sequential Recommendation} 

\begin{figure*}[t!]
	\centering
	\includegraphics[width=0.8\textwidth]{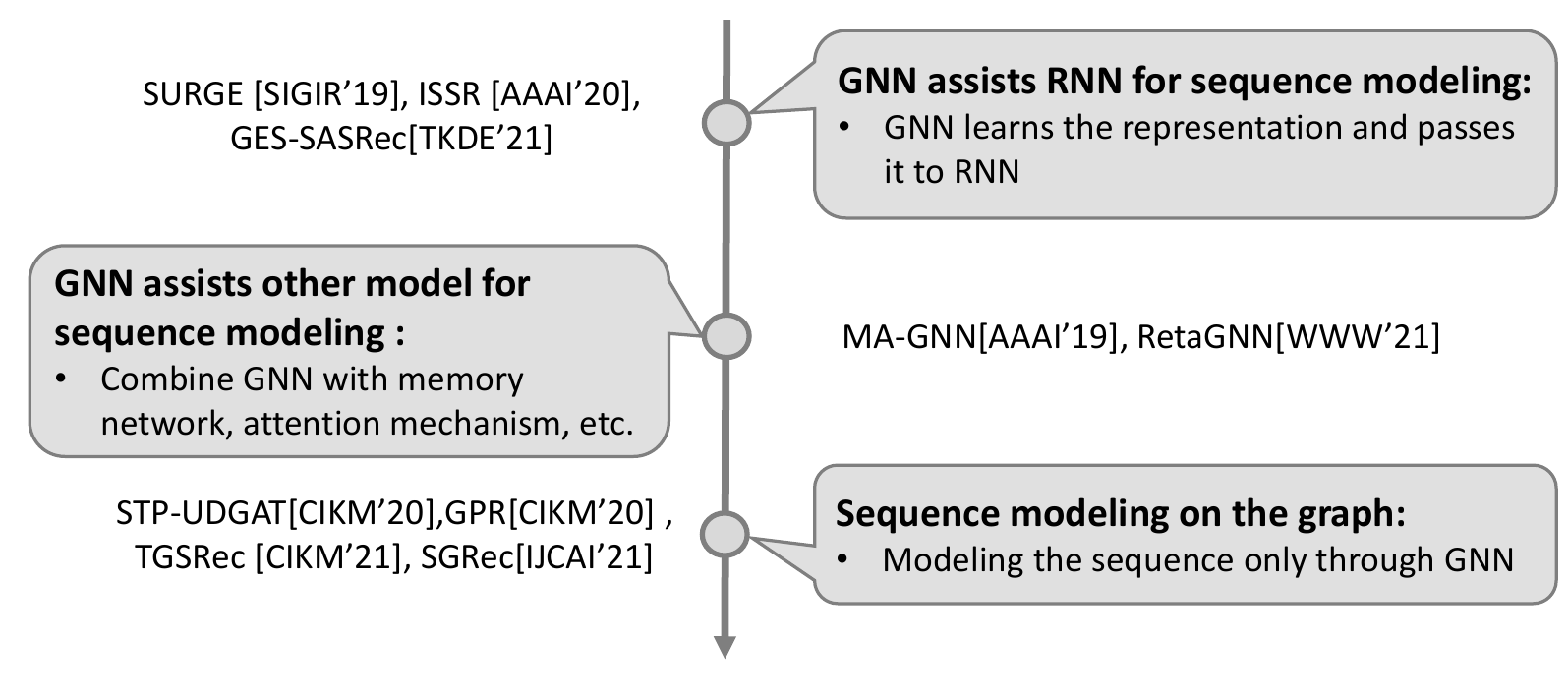} 
	\caption{Illustration of GNN models for sequential recommendation.} \label{fig:sequence_modeling}
\end{figure*}

For sequential recommendation, in order to improve the recommendation performance, it is necessary to extract as much effective information as possible from the sequence, and to learn the user's interest in the sequence, including short-term interest, long-term interest, dynamic interest, etc., to accurately predict the next item that the user may be interested in. Some tools for sequence modeling have been used, such as Markov chains~\cite{cheng2013you} or recurrent neural networks~\cite{kang2018self}. Graph neural networks can be well leveraged for short-term, dynamic interest modeling or representation learning by converting the data to a graph.
A general pattern for sequential modeling with GNN is shown in figure~\ref{fig:sequence_modeling}.

SURGE~\cite{chang2021sequential} transforms the sequence of each user into an item-item graph and adaptively learns the weights of edges through metric learning, with only the stronger edges retained by dynamic graph pooling. The retained graph is converted to a sequence by position flattening and finally be used to predict the next item. Ma \textit{et al.} ~\cite{ma2020memory} considers the short-term interest modeling in the sequence to build an item-item graph. For each item, this work only builds edges with other items close to it in the sequence. This enables it to learn short-term user interests in the sequence while still learning long-term user interests through other networks. The learned multiple representations are fused together and used for final recommendation.

\begin{table*}[]
	\centering
	\small
	\caption{Details of GNN models for sequential recommendation.}\label{tab:sequential}
	\begin{tabular}{cccc}
		\hline
		Model &  \revise{Graph Construction}  & \revise{Network Design} & Sequential Modeling   \\
		\hline
		SURGE \cite{chang2021sequential}  & item-item graph &  GAT & RNN \\
		\cite{wang2020knowledge}  & item-item graph & GCN & Attention \\
		ISSR\cite{liu2020inter}  & item-item and user-item graph & GCN & RNN\\
		MA-GNN\cite{ma2020memory}  &  item-item graph & GCN & Memory network\\
		DGSR\cite{zhang2021dynamic} & user-item graph & GAT & RNN \\
		GES-SASRec\cite{zhu2021graph} & item-item graph & GCN & RNN \\
		RetaGNN\cite{hsu2021retagnn} & temporal heterogeneous graph & GAT & Self Attention \\
		TGSRec\cite{fan2021continuous} & temporal user-item graph & GAT & GAT \\
		SGRec\cite{li2021discovering} & item-item graph & GAT & GAT \\
		GME\cite{xie2016graph} & item-item and user-item graph & GAT & GAT \\
		STP-UDGAT\cite{lim2020stp} & item-item and user-item graph& GAT & GAT \\
		GPR\cite{chang2020learning} & item-item and user-item graph & GCN & GCN \\
		
		\hline
	\end{tabular}
\end{table*}

\revise{Besides the challenges of graph construction and network design, the methods of GNN-based sequential recommendation should carefully design how to extract the sequential patterns in users’ behaviors, as shown in Table~\ref{tab:sequential}.}
Since GNN has the ability of high-order relationship modeling by aggregating information from neighbor nodes, after fusing multiple sequences into one graph, it can learn representations of both users and items in different sequences, which can't be accomplished by the Markov model or recurrent neural network. Wang \textit{et al.}~\cite{wang2020knowledge} propose a simple method that directly converts the sequence information into directed edges on the graph and then uses GNN to learn representations. Liu \textit{et al.} ~\cite{liu2020inter} construct a user-item bipartite graph and an item-item graph at the same time, where the edges of the item-item graph indicate co-occurrence in a sequence, with edge weights assigned according to the number of occurrences. The representations learned by GNN are used in the final recommendation through the recurrent neural network. Different from directly converting the temporal sequence into directed edges in the graph, DGSR~\cite{zhang2021dynamic} and TGSRec~\cite{fan2021continuous} consider the timestamps in the sequence in the process of graph construction. In the graph, each edge represents the interaction between the user and the item, along with the corresponding time attribute. Then perform convolution operations on the temporal graph to learn the representations of users and items. GES-SASRec~\cite{zhu2021graph} and SGRec~\cite{li2021discovering} focus on the learning of item representations. For an item in a sequence, GES-SASRec~\cite{zhu2021graph} considers the next item of this item in other sequences, and SGRec~\cite{li2021discovering} not only considers the next item but also considers the previous one. By aggregating the items before and after the target item in different sequences, the representation of the item is enhanced. GPR~\cite{chang2020learning} and GME~\cite{xie2016graph} construct edges between items by considering the frequency of consecutive occurrences or occurrences in the same sequence to enhance the representation. Some works are more complicated. For example, RetaGNN~\cite{hsu2021retagnn} considers the attributes of the items when constructing the graph, while STP-UDGAT~\cite{lim2020stp} considers the geographic location, timestamp, and frequency in the POI recommendation. Table~\ref{tab:sequential} summarizes the above works.
\revise{Current networks of graph neural network-based recommendation are still highly relying on an additional sequence model such as RNN, Transformer, etc., for sequential modeling. The ability of graphs in sequential modeling, such as directed edges, dynamic nodes, requires more research efforts.}

\subsubsection{GNN in Session-based Recommendation }
\begin{figure}[t!]
	\centering
	\includegraphics[width=0.8\textwidth]{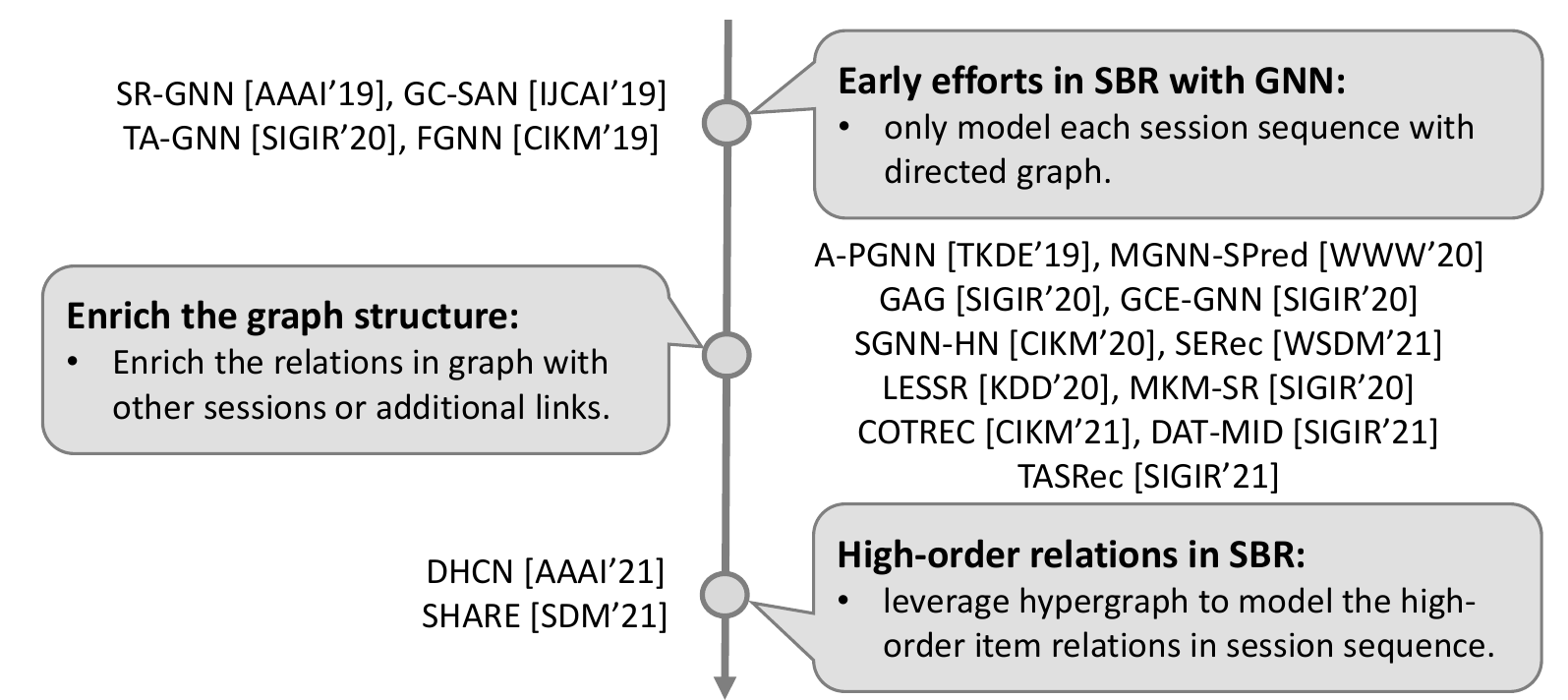} 
	\caption{Illustration of GNN models for session-based recommendation (SBR).} \label{fig:session_map}
\end{figure}

\begin{table*}[t!]
	\centering
	\small
	\caption{Details of GNN models for session-based recommendation.}\label{tab:SBR}
	\begin{tabular}{cccc}
		\hline
		Model & \revise{Graph Construction}   & \revise{Network Design}  & Enrich Graph Structure\\
		\hline
		SR-GNN~\cite{SRGNN}  & directed graph  & gated GNN  & -\\
		GC-SAN~\cite{xu2019GC-SAN}  & directed graph  & gated GNN  & -\\
		TA-GNN~\cite{yu2020tagnn} & directed graph  & gated GNN  & -\\
		FGNN~\cite{FGNN} & directed graph  & GAT  & -\\
		A-PGNN~\cite{wu2019A-PGNN} & directed graph  & gated GNN  & cross sessions\\
		MGNN-SPred~\cite{wang2020MGNN-SPred} & directed \& multi-relational item graph & GraphSAGE  & cross sessions \\
		GAG~\cite{qiu2020GAG} & directed graph & GCN  & cross sessions\\
		SGNN-HN~\cite{pan2020SGNN-HN} & star graph  & gated GNN  & additional edges \\
		GCE-GNN~\cite{GCEGNN} & directed graph + global graph & GAT & cross sessions \\
		DGTN~\cite{zheng2020DGTN} & directed graph & GCN & cross sessions \\
		DHCN~\cite{DHCN} & hypergraph + line graph & HyperGCN & cross sessions \\
		SHARE~\cite{SHARE} & session hypergraph & HyperGAT & additional edges \\
		SERec~\cite{chen2021SERec} & KG + directed graph & GAT + gated GNN &  cross sessions \\ 
		LESSR~\cite{chen2020LESSR} & directed graph & GAT & additional edges\\ 
		CAGE~\cite{sheu2020CAGE}  & KG + article-level graph & GCN & cross sessions\\ 
		MKM-SR~\cite{meng2020MKM-SR}  & KG + directed graph & gated GNN & cross sessions \\ 
		COTREC~\cite{xia2021COTREC}  & item graph + line graph & GCN &  cross sessions\\ 
		DAT-MDI~\cite{chen2021DAT-MDI} & directed graph + global graph & GAT & cross sessions \\  
		TASRec~\cite{zhou2021TASRec} & dynamic graph & GCN & cross sessions \\ 
		
		HG-GNN~\cite{pang2022heterogeneous} &directed graph+global graph &gated GNN& cross sessions\\
		CGL~\cite{pan2022collaborative} &directed graph+global graph&gated GNN&cross sessions \\
		
		\hline
	\end{tabular}
\end{table*}
In session-based recommendation, the session data may contain both user interests and noisy signals. 
Suppose a session for a certain user, \textit{iPhone} $\rightarrow$ \textit{iPad} $\rightarrow$ \textit{milk} $\rightarrow$ \textit{AirPods}. Obviously, \textit{milk} is likely clicked by mistake and then becomes a noise, as the session reflects the user’s preference for electronic products.
Hence, the two main considerations in the session-based recommendation are 1) how to model the item transition pattern in session data, and 2) how to activate the user's core interests from noisy data.
From the perspective of graph learning, the item transitions can be modeled as graph, and the information propagation on the graph can activate the user's actual interests.
\begin{itemize}[leftmargin=*]
	\item \textbf{Graph construction.} In session-based recommendation, most existing works~\cite{SRGNN,FGNN,GCEGNN,xu2019GC-SAN,yu2020tagnn} model the session data with a directed graph to capture the item transition pattern. Distinct from sequential recommendation, the session sequence in session-based recommendation is short and the user behaviors are limited, \textit{i.e.}, the average length of sequences in Tmall\footnote{\url{https://www.tmall.com}} is only 6.69~\cite{GCEGNN,DHCN}. Hence, a session graph constructed from a single session may only contain limited nodes and edges. To address the above challenge and sufficiently capture the possible relations among items, there are two strategies, 1) straightforwardly capturing relations from other sessions and 2) adding the additional edges of the session graph. For the first strategy, A-PGNN~\cite{wu2019A-PGNN}, DGTN~\cite{zheng2020DGTN}, and GAG~\cite{qiu2020GAG} propose to enhance relations of the current session graph with related sessions, and GCE-GNN~\cite{GCEGNN} leverages the global context by constructing another global graph to assist the transition patterns in the current session. DHCN~\cite{DHCN} regards each session as a hyperedge and represents all sessions in a hypergraph to model the high-order item relations. SERec~\cite{chen2021SERec} enhances the global information for each session with a knowledge graph. 
	CAGE~\cite{sheu2020CAGE} learns the representations of semantic-level entities by leveraging the open knowledge graph to improve the session-based news recommendation.
	MKM-SR~\cite{meng2020MKM-SR} enhances the information in the given session by incorporating user micro-behaviors and item knowledge graph. COTREC~\cite{xia2021COTREC} unifies all sessions into a global item graph from the item view and captures the relations among sessions by line graph from the session view. DAT-MID~\cite{chen2021DAT-MDI} follows GCE-GNN~\cite{GCEGNN} to construct both session graph and global graph and then learns item embeddings from different domains.
	TASRec~\cite{zhou2021TASRec} constructs a graph for each day to model the relations among items and enhance the information in each session.
	As for the second strategy, SGNN-HN~\cite{pan2020SGNN-HN} constructs a star graph with a "star" node to gain extra knowledge in session data. SHARE~\cite{SHARE} expands the hyperedge connections by sliding the contextual window on the session sequence. LESSR~\cite{chen2020LESSR} proposes first to construct an edge-order preserving multigraph and then construct a shortcut graph for each session for enriching edge links.
	A$^{3}$SR~\cite{deng2022g} proposes to construct one global graph to represent all sessions, and distinguish session-level prediction signals based on the later embedding propagation.
	HG-GNN~\cite{pang2022heterogeneous} proposed to construct a heterogeneous graph in which the item transitions are captured by two types of item-item edge: in-type and out-type.
	CGL~\cite{pan2022collaborative}  also proposed to construct a global-level graph and a session-level graph, but it considers the learning on two graphs as two tasks, based on which multi-task learning is adopted.

	\item \textbf{Information propagation.} 
	As for the information propagation on the constructed graph, there are four propagation mechanisms that are used in session-based recommendation, e.g., gated GNN, GCN, GAT, and GraphSAGE. SR-GNN~\cite{SRGNN} and its related works~\cite{xu2019GC-SAN,yu2020tagnn,pan2020SGNN-HN,wu2019A-PGNN,meng2020MKM-SR,chen2021SERec} combine the gated recurrent units in the propagation (gated GNN) on the session graph. GAG~\cite{qiu2020GAG}, DCTN~\cite{zheng2020DGTN} conduct graph convolution on the constructed directed graph.
	DHCN~\cite{DHCN} proposes to perform graph convolution on both hypergraph and line graph to obtain session representations from two different perspectives.
	Similar to DHCN~\cite{DHCN}, COTREC~\cite{xia2021COTREC} performs GCN on item graph and line graph to obtain information from item and session views, respectively.
	CAGE~\cite{sheu2020CAGE} conducts GCN on article-level graph and TASRec~\cite{zhou2021TASRec} performs graph convolution on the dynamic graph to capture the item relations.
	FGNN~\cite{FGNN} conducts GAT on a directed session graph to assign different weights to different items. SHARE~\cite{SHARE} performs GAT on session hypergraphs to capture the high-order contextual relations among items.
	GCE-GNN~\cite{GCEGNN} and DAT-MID~\cite{chen2021DAT-MDI} perform GAT on both the session graph and global graph to capture the local and global information, respectively.
	MGNN-SPred~\cite{wang2020MGNN-SPred} adopts GraphSAGE on a multi-relational item graph to capture the information from different types of neighbors.
	
\end{itemize}

Table \ref{tab:SBR} shows the differences among the above approaches for session-based recommendation. 
To sum up, the development of session-based recommendation (SBR) with GNN can be summarized in Fig.~\ref{fig:session_map}. Early efforts in SBR only model each session sequence with a directed graph, such as SR-GNN~\cite{SRGNN}, GC-SAN~\cite{xu2019GC-SAN}, TA-GNN~\cite{yu2020tagnn}, and FGNN~\cite{FGNN}. Then, some methods attempt to enrich the relation and information in the session graph with other sessions or additional links. The methods in~\cite{wu2019A-PGNN,wang2020MGNN-SPred,qiu2020GAG,GCEGNN,zheng2020DGTN,DHCN,chen2021SERec,sheu2020CAGE,meng2020MKM-SR,xia2021COTREC,chen2021DAT-MDI,zhou2021TASRec} combine the information from other sessions to capture more information from similar sessions or all sessions. 
Moreover, the methods with additional links, such as SGNN-HN~\cite{pan2020SGNN-HN}, SHARE~\cite{SHARE} and LESSR~\cite{chen2020LESSR}, attempt to introduce additional edges to the given session to capture the complex relations and information in session data.
Furthermore, some studies, such as DHCN~\cite{DHCN} and SHARE~\cite{SHARE}, attempt to enhance the recommendation by modeling the high-order relations in session data more sufficiently. 
\revise{As we can observe, the current works of GNN models for session-based recommendation have been developed from simple graphs to complex graphs, with even hyper-edges. Unlike sequential recommendation, session-based recommendation, in which users have repeated interactions with specific items, is more suitable for GNN-based models. This makes the existing works of session-based recommendation can build more powerful and complex graph structures. A possible improvement of this area is to explore larger-scale datasets and further evaluate GNN's ability in these datasets.}

\subsubsection{GNN in Bundle Recommendation}

\begin{figure}[t!]
	\centering
	\includegraphics[width=0.8\textwidth]{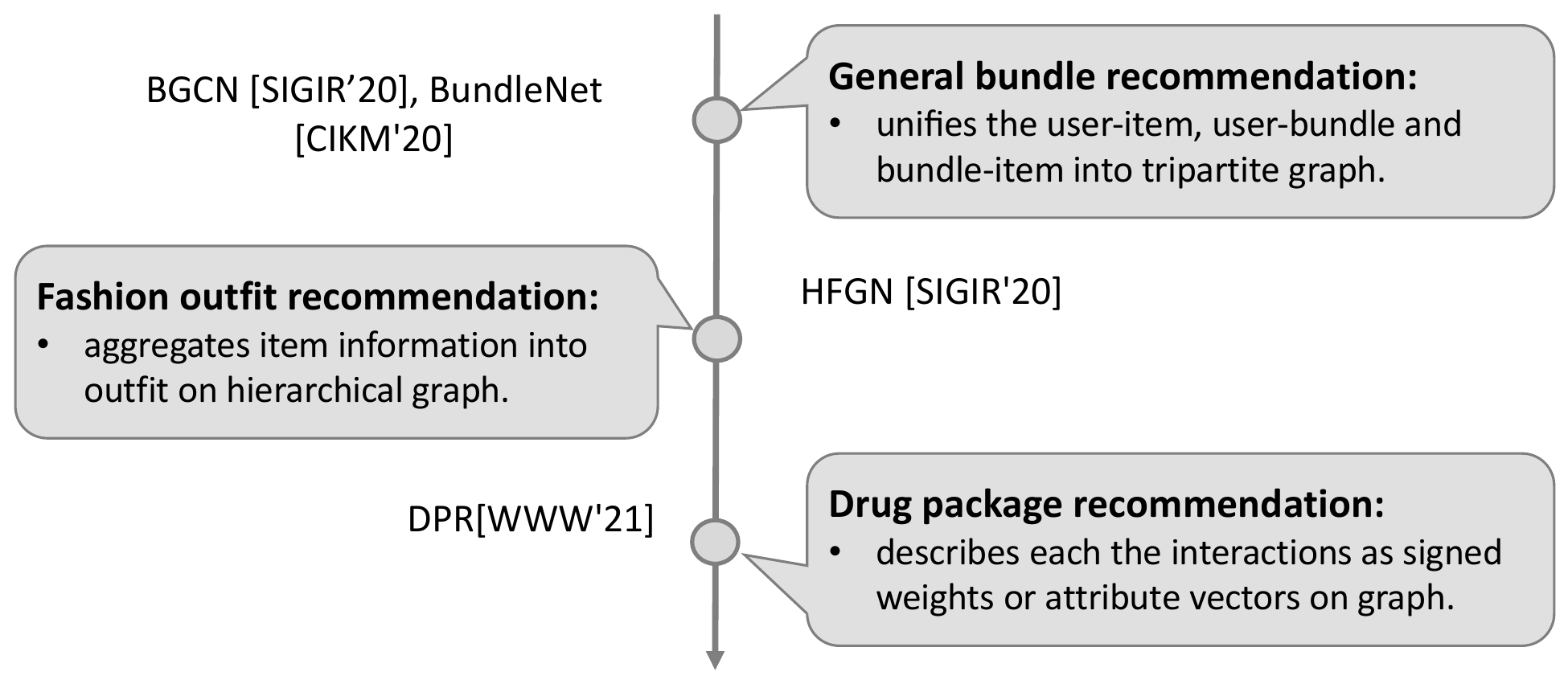} 
	\caption{Illustration of GNN models for bundle recommendation.} \label{fig:bundle_rec_work}
\end{figure}

\begin{table*}[t!]
	\footnotesize
	\centering
	\caption{Details of GNN models for bundle recommendation.}\label{tab:bundle}
		\begin{tabular}{cccc}
			\hline
			Model & \revise{Graph construction for affiliation} & \revise{Refining bundle} & \revise{Network design for High-order Relation} \\
			\hline
			BGCN~\cite{BGCN} & tripartite graph & meta-path propagation & multi-layer GCN \\
			HFGN~\cite{HFGN} & hierarchical Graph & direct aggregation & multi-layer GCN \\
			BundleNet~\cite{BundleNet} & tripartite graph & transformation parameters & multi-layer GCN \\
			DPR~\cite{DPR} & tripartite graph & graph induction & multi-layer GCN\\
			DPG~\cite{zheng2022interaction} & tripartite graph & graph induction & multi-layer GCN\\
			MIDGN~\cite{zhao2022multi} & tripartite graph &graph induction &multi-layer GCN \& disentangling\\
			\hline
		\end{tabular}
	\end{table*}
	
	The three challenges of bundle recommendation are 1) users' decisions towards bundles are determined by the items the bundles contain (the affiliation relation), 2) learning bundle representations with the sparse user-bundle interactions and 3) high-order relations.
	The earlier works approach the problem of bundle recommendation by learning from user-item interaction and user-bundle interaction together, with parameter sharing or joint loss function~\cite{BBPR, DAM}.
	For the first time, Chang \textit{et al.}~\cite{BGCN} propose a GNN-based model that unifies both two parts of interactions and the bundle-item affiliation-relations into one graph.
	Then the item can serve as the bridge for embedding propagation between user-bundle and bundle-bundle. Besides, a specially designed sampling manner for finding hard-negative samples is further proposed for training.
	Deng \textit{et al.}~\cite{BundleNet} construct a similar tripartite graph with the transformation parameters to well extract bundle representations from included items' representations.
	Zheng \textit{et al.}~\cite{DPR} consider the bundle recommendation in the drug package dataset and propose to initialize a graph with auxiliary data and represent the interaction as scalar weights and vectors. GNN layers are proposed for obtaining the drug package embeddings.
	This method is further improved by~\cite{zheng2022interaction} by modeling the interaction data with the graph consisting of all bundles to obtain bundle embeddings and introducing a goal of new-bundle generation based on reinforcement learning.
	Li \textit{et al.}~\cite{HFGN} consider the problem of personalized outfit recommendation, which can also be regarded as a kind of bundle recommendation. The authors construct a hierarchical graph in which the users, items, and outfits are contained.
	GNN layers are deployed for obtaining the representations of users and outfits. The learning of the GNN model follows a multi-task manner.
	Zhao~\textit{et al.}~\cite{zhao2022multi} proposed a MIDGN model which combines the disentangle embedding space and multi-layer embedding propagation, obtaining finer-grained user preference towards items and bundles.
	
	In summary, the data input of bundle recommendation can be well represented as graph-structural data, especially for the bundles, which have been represented as a kind of new nodes.
	
	\subsubsection{GNN in Cross-Domain Recommendation}

	\begin{figure}[t!]
		\centering
		\includegraphics[width=0.8\textwidth]{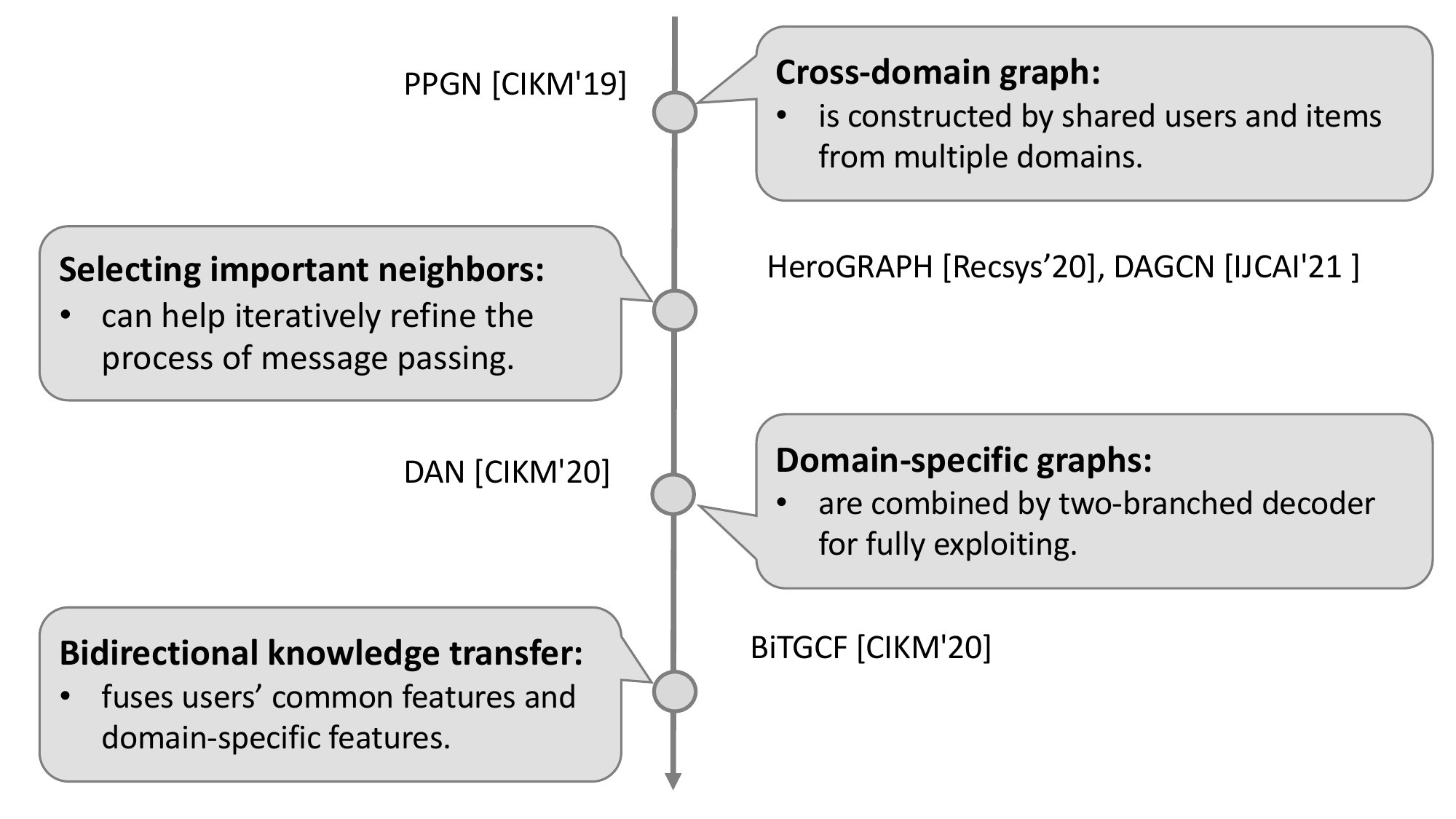} 
		\caption{Illustration of GNN models for cross-domain recommendation.} \label{fig:cross_domain_rec_work}
	\end{figure}
	\begin{table*}[t!]
		\centering
		\small
		\caption{Details of GNN models of cross-domain recommendation.}\label{tab:cross}
		\begin{tabular}{cccc}
			\hline
			Model & \revise{Graph Construction} & \revise{Network design for Information Transferring} \\
			\hline
			PPGN\cite{zhao2019cross} & cross-domain graph & cross-domain propagation \\
			BiTGCF\cite{BiTGCF} & domain-specific graphs & common user attributes \\
			DAN\cite{DAN} & domain-specific graphs & two-branched decoder \\
			HeroGRAPH\cite{HeroGRAPH} & cross-domain graph & cross-domain propagation \\
			DAGCN\cite{DAGCN} & cross-domain graph & cross-domain propagation \\
			\hline
		\end{tabular}
	\end{table*}
	
	Benefitting from the powerful capabilities, the GNN-based recommendation model has gradually emerged in the cross-domain recommendation. 
	\revise{The challenges include how to construct graph and design network architecture to transfer information across domains, as shown in Table \ref{tab:cross}.}
	Zhao \textit{et al.}~\cite{zhao2019cross} construct the cross-domain graph with shared users and items from multiple domains. The proposed PPGN's embedding propagation layers can well learn the users' preferences on multiple domains' items under a multi-task learning framework.
	Liu \textit{et al.}~\cite{BiTGCF} propose the bidirectional knowledge transfer by regarding the shared users as the bridge. GNN layers are adopted to leverage the high-order connectivity in the user-item interaction graph for better preference learning,
	and then it fuses common features with domain-specific features.
	Guo \textit{et al.}~\cite{DAGCN} construct the graph for each domain and deploy domain-specific GCN layers for learning user-specific embeddings. The authors combine it with the attention mechanisms for adoptively choosing important neighbors during the embedding propagation.
	Wang \textit{et al.}~\cite{DAN} propose an encoder-decoder framework where the encoder is implemented by graph convolutional networks. Specifically, the GCNs are deployed on the user-item interaction graphs. 
	Cui \textit{et al.}~\cite{HeroGRAPH} proposes to construct a heterogeneous graph, where users and items in multiple domains can be well included. The GNN-based embedding propagation is deployed on multiple domains, where a user/item can directly absorb the information of different domains, where recurrent attention networks are used to distinguish important neighbors.
	\revise{To sum up, the cross-domain recommendation is a general definition, which covers various specific settings. For example, one setting may assume that the user can be overlapped while another setting may be on the contrary. Graph neural networks have well studied a few, leaving others important settings as future works.}
	
	\subsubsection{GNN in Multi-behavior Recommendation}
	
	\begin{figure}[t!]
		\centering
		\includegraphics[width=0.8\textwidth]{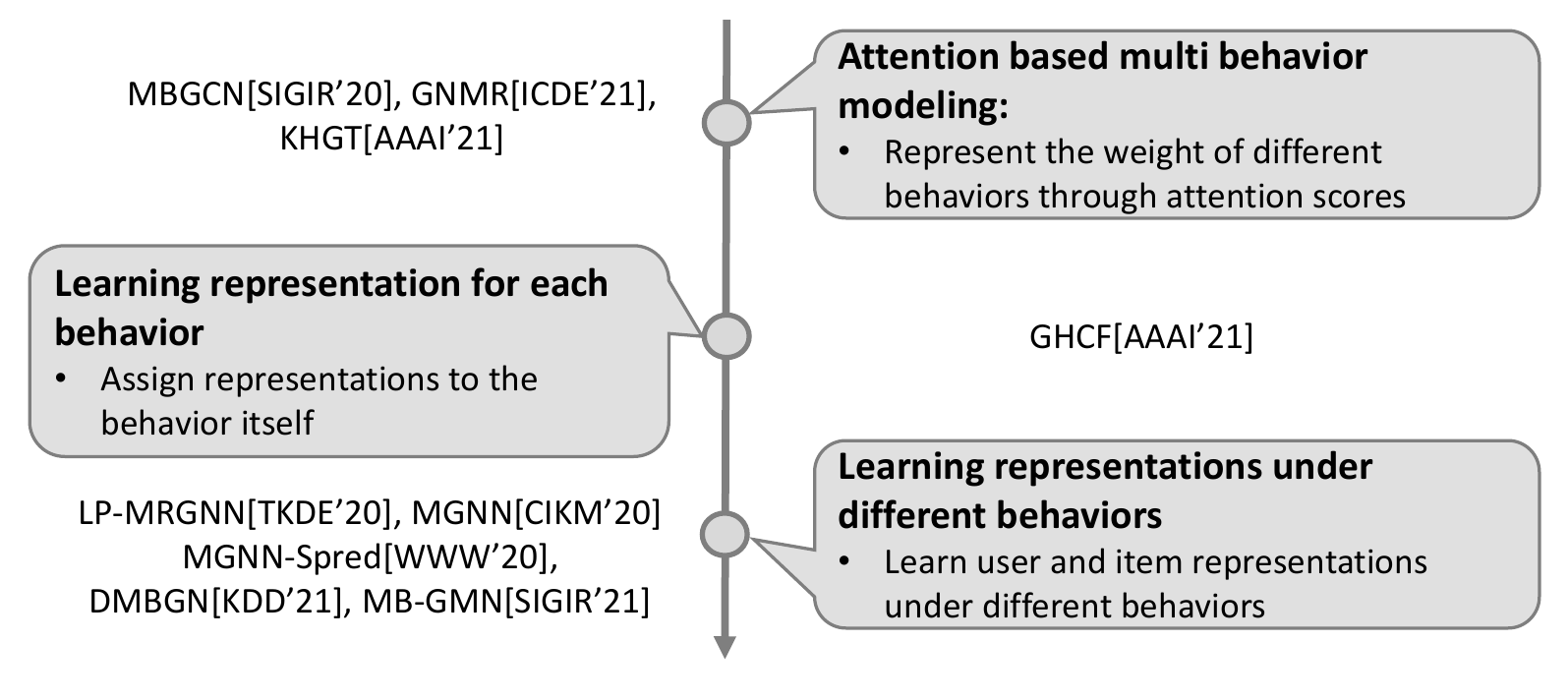} 
		\caption{Illustration of GNN models for multi-behavior recommendation.}
		\label{fig:gnn_multi_behavior}
	\end{figure}
	
	\begin{table*}[t!]
		\centering
		\small
		\caption{Details of GNN models for multi-behavior recommendation.}\label{tab:multi_behavior}
		\begin{tabular}{cccc}
			\hline
			Model & \revise{Graph Construction} & \revise{Network Design} & Multi-Behavior Modeling \\
			\hline
			
			MB-GMN\cite{xia2021graph}  & user-item graph & GCN / GAT & Behavior representation \\
			KHGT\cite{xia2021knowledge}  & user-item / item-item graph & GAT & Weight \\
			MBGCN\cite{jin2020multi}  & user-item graph & GAT & Weight \\
			MGNN-SPred\cite{wang2020MGNN-SPred} & item-item graph & GCN & Weight \\
			MGNN\cite{zhang2020multiplex}  & user-item graph & GCN  & Node representation\\
			LP-MRGNN\cite{wang2021incorporating} & item-item graph & GCN & Weight \\
			GNNH\cite{yu2021graph} & item-item / category-category graph & GCN & Item representation \\ 
			GNMR\cite{xia2021multi} & user-item graph & GAT & Weight \\
			DMBGN\cite{xiao2021dmbgn} & item-item graph & GCN & Graph representation \\
			GHCF\cite{chen2021graph}  & user-item graph & GCN & Edge embedding\\
			\hline
		\end{tabular}
	\end{table*}

	Multiple types of behaviors can provide a large amount of information to the recommender system, which helps the recommender system to learn the user's intentions better, thereby improving the recommendation performance. For recommender systems based on graph neural networks, based on standard user-item bipartite graphs, the multiple types of behaviors between users and items can naturally be modeled as different types of edges between nodes. Therefore, most of the multi-behavior recommendation methods based on graph neural networks are based on heterogeneous graphs. However, the focus of multi-behavior recommendation is 1) how to model the relationship between multiple behaviors and target behavior, and 2) how to model the semantics of the item through behavior, which is shown in Fig.~\ref{fig:gnn_multi_behavior}. 
	
	In order to model the effect of auxiliary behaviors on target behaviors, the simplest method is to directly model all types of behaviors without considering the differences between behaviors. Zhang \textit{et al.}~\cite{zhang2020multiplex} constructs all user behaviors in one graph and performs graph convolution operations. Wang ~\cite{wang2020MGNN-SPred, wang2021incorporating} extracts each behavior from the graph to construct a subgraph, then learns from the subgraph, and finally aggregates through the gating mechanism. Chen \textit{et al.} ~\cite{chen2021graph} proposed a intuitive method which assigns a representation to users, items, and behaviors. In the propagation process, the representations of edges and neighbor nodes need to be composited first to obtain a new node representation, and then it can be applied to the GNN method. Through the composition operation, the representation of the node is fused with different types of behaviors.
	Xia \textit{et al.} ~\cite{xia2021multi} also redesign the aggregation mechanism on the graph convolutional network to explicitly model the impact of different types of behavior. Jin \textit{et al.} ~\cite{jin2020multi} assign different learnable weights to different edges to model the importance of the behaviors. In addition, in order to capture complex multi-behavior relationships, some works rely on knowledge. For example, Xia \textit{et al.} ~\cite{xia2021knowledge} learns the representations in different behavior spaces and then injects temporal contextual information into the representations to model the user's behavior dynamics, and finally, through the attention mechanism, discriminate the most important relationships and behaviors for the predicted target. Xia \textit{et al.} ~\cite{xia2021graph} uses meta-graph networks to learn meta knowledge for different behaviors and then transfer the learned meta-knowledge between different types of behaviors.

	In addition to modeling the influence of different types of behaviors, different behaviors may contain different meanings or semantics. For example, for items added to a shopping cart, users may have similar preferences for them, or these items have a complementary relationship and generally need to be purchased at the same time. If these items are connected through a graph, the representation of the item can be enhanced. 
	In order to get better item representation, Yu \textit{et al.}~\cite{yu2021graph} not only connects related items in the graph but also constructs a new graph of the category that the items belong to, which is used to enhance the representation of the item. It needs to construct the item graph separately, and there are also some works that do not construct the graph separately and directly in the user-item heterogeneous graph by using meta-path or second-order neighbors, where similar items are aggregated to enhance their representation~\cite{jin2020multi, xia2021knowledge}. These works' details are presented in Table~\ref{tab:multi_behavior}.
	
	\subsection{GNN for Different Recommendation Objectives}

	\subsubsection{GNN for Diversity}
	For individual-level diversity, retrieved items from recommender systems need to cover more topics such as different categories of products or different genres of music.
	Therefore, utilizing GNN to increase diversity requires that the learned user embeddings be close to item embeddings with various topics.
	However, as the embedding aggregation operation in GNN makes user embeddings close to embeddings of items that are interacted in historical records, GNN might discourage diversity by recommending too many similar items that belong to the dominant topic in users' interaction history.
	For example, the learned embedding from GNN for a user who mainly interacts with electronics may be too close to embeddings of electronic items, making the GNN only recommend electronics to this user, which leads to low diversity.
	Therefore, to overcome the first challenge of weak signals from disadvantaged topics, efforts have been made to restrict the importance of dominant topics (electronics in the above example) by constructing diversified sub-graphs from the original user-item bipartite graph.
	Specifically, Sun \textit{et al.} \cite{sun2020framework} propose a model called Bayesian Graph Collaborative Filtering (BGCF), which constructs augmented graphs with node copying \cite{pal2019bayesian} from high-order neighbors, such that items of diverse topics with high similarity can be directly connected to user nodes.
	Zheng \textit{et al.} \cite{zheng2021dgcn} propose Diversified Graph Convolutional Networks (DGCN) and conduct rebalanced neighbor sampling, which down-weights dominant topics and boosts the importance of disadvantaged topics in neighbor nodes.
	Fig. \ref{fig:diversity_model} illustrates the comparison between BGCF and DGCN on graph construction for diversity.
	Meanwhile, to address the second challenge of balancing between accuracy and diversity, BGCF re-ranks the top items according to item popularity, and DGCN utilizes adversarial learning on the item embeddings which encourages GNN to capture user preferences that are largely independent of item categories, which further improves recommendation diversity.

	\begin{figure}[t]
		\centering
		\includegraphics[width=0.9\textwidth]{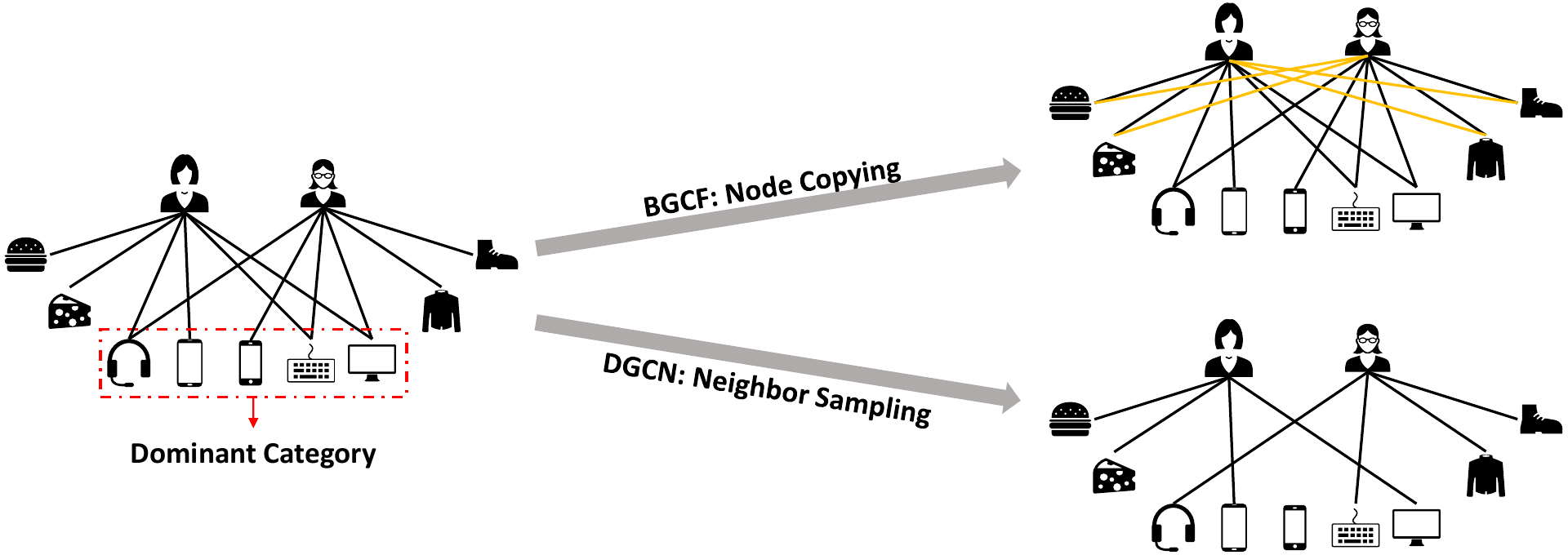} 
		\caption{Illustration of BGCF \cite{sun2020framework} and DGCN \cite{zheng2021dgcn} on achieving individual-level diversity.} \label{fig:diversity_model}
	\end{figure}
	
	\begin{figure}[t]
		\centering
		\includegraphics[width=0.7\textwidth]{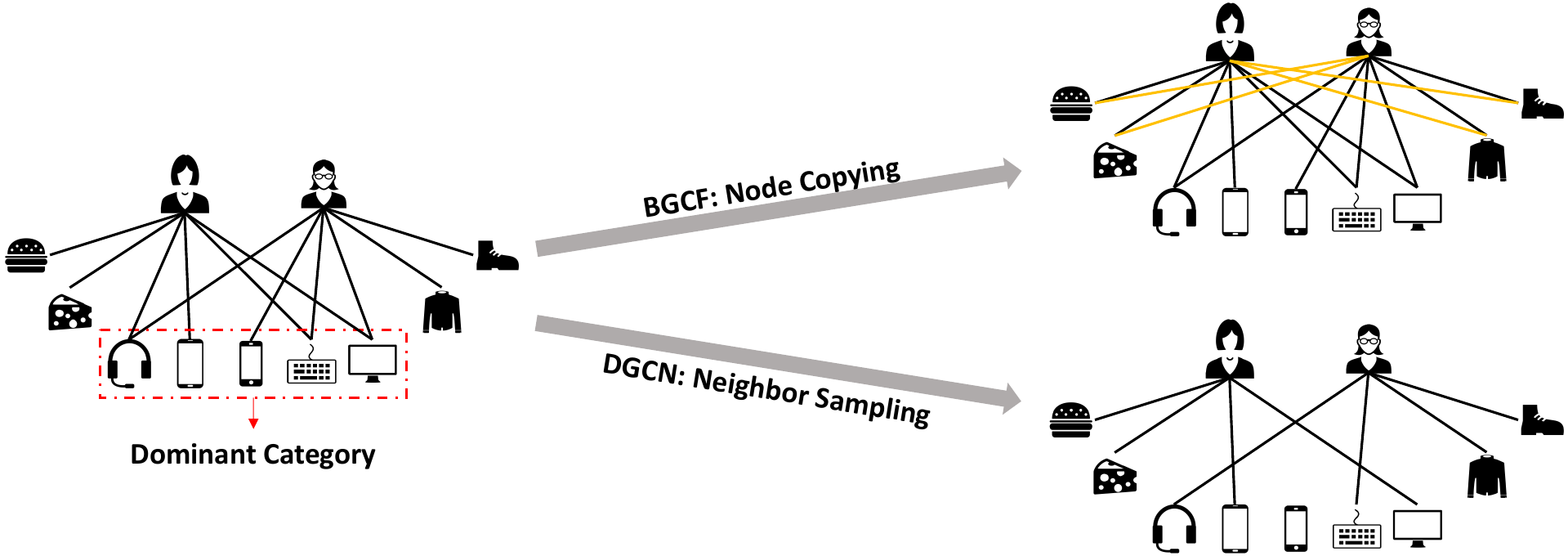} 
		\caption{\revise{Illustration of V2HT~\cite{li2019long}, FH-HAT~\cite{xie2021improving}, and Tradeoff-Framework~\cite{isufi2021accuracy} on achieving system-level diversity.}} \label{fig:diversity_model}
	\end{figure}
	
	\begin{table}[t!]
		\centering
		\small
		\caption{Details of GNN models for diversified recommendation (how they address challenges)}\label{tab:diversity}
		\begin{tabular}{ccc}
			\hline
			Model & Weak Preference Signals & Accuracy-Diversity Balance \\
			\hline
			BGCF \cite{sun2020framework}  & Node Copying  & Re-rank          \\
			DGCN \cite{zheng2021dgcn} & Neighbor Sampling & Adversarial training \\
			V2HT \cite{li2019long}     & Utilize item correlations       & - \\
			FH-HAT \cite{xie2021improving}  & Heterogeneous graph & Diversity loss \\
			Tradeoff-Framework\cite{isufi2021accuracy} & NN \& FN graph & Joint training \\
			\hline
		\end{tabular}
	\end{table}
	As for system-level diversity, the main target is to discover more relevant items from the long-tail ones, which have much fewer training samples than those popular items.
	To address the weak signals of long-tail items, Li \textit{et al.} \cite{li2019long} propose a model called V2HT to construct an item graph that explores item correlations with external knowledge.
	Specifically, four types of edges are introduced, which connect frequent items and long-tail items.
	Then multiple GCN layers are stacked, which propagates well-trained embeddings of frequent items to undertrained embeddings of long-tail items.
	In this way, long-tail item embeddings of higher quality are obtained since they share the information from frequent items; thus, system-level diversity is improved with more recommendation on long-tail items.

	In addition, a few studies \cite{xie2021improving,isufi2021accuracy} utilize GNN to improve both individual-level and system-level diversity.
	Specifically, Xie \textit{et al.} \cite{xie2021improving} propose FH-GAT, which addresses the challenge of weak signals by constructing a heterogeneous interaction graph to express diverse user preferences.
	A neighbor similarity-based loss is conducted on the heterogeneous graph to balance accuracy and diversity.
	Isufi \textit{et al.} \cite{isufi2021accuracy} propose two GCN on the nearest neighbor (NN) graph and the furthest neighbor (FN) graph, where NN guarantees accuracy and FN enhances the weak signals of diverse items.
	Meanwhile, the two GCN are jointly optimized with a hyper-parameter to achieve a trade-off between accuracy and diversity.
	Table \ref{tab:diversity} shows the differences among the above approaches.

	\subsubsection{GNN for Explainability}

	\begin{figure}[t!]
		\centering
		\includegraphics[width=0.70\textwidth]{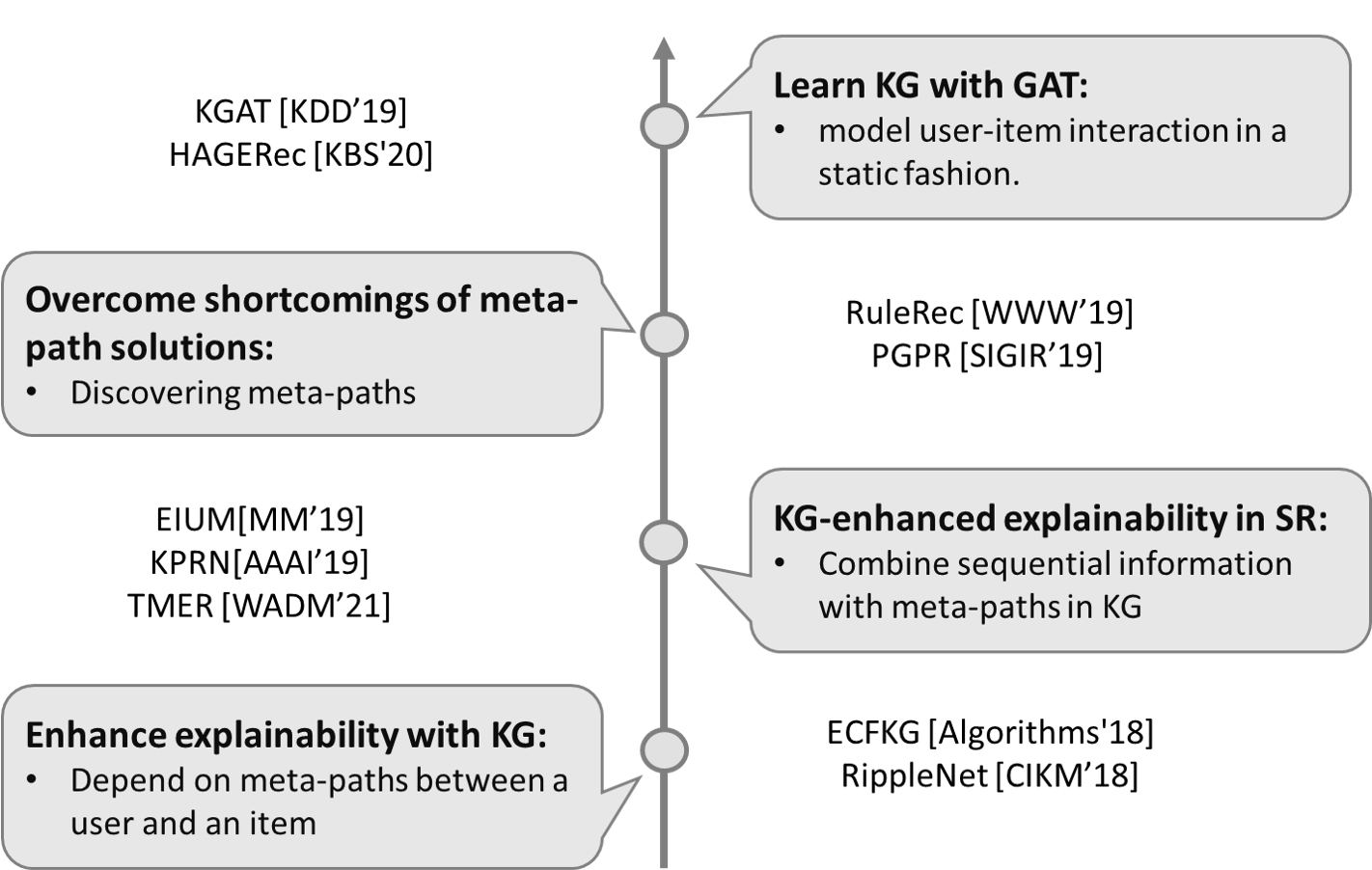} 
		\caption{Illusration of GNN models for explainable recommendation} \label{fig:explainability}
	\end{figure}

	\begin{table}[t!]
		\centering
		\small
		\caption{Details of GNN models for explainable recommendation}\label{tab:explainability}
		\begin{tabular}{ccc}
			\hline
			Model & \revise{Network Design}  & Source of Explainability \\
			\hline
			ECFKG~\cite{ai2018learning}  & Meta-path based GE & Meta-path over knowledge graph \\
			RippleNet~\cite{wang2018ripplenet} & Meta-path based GraphSage & Meta-path over knowledge graph \\
			EIUM~\cite{huang2019explainable} & Meta-path based GE & Meta-path over knowledge graph \\
			KPRN~\cite{wang2019explainable} & Meta-path based GE & Meta-path over knowledge graph \\
			TMER~\cite{chen2021temporal} & Meta-path based GE & Meta-path over knowledge graph and temporal dependency \\
			RuleRec~\cite{ma2019jointly} & Meta-path based GE & Discovered meta-path over knowledge graph \\
			PGPR~\cite{Xian2019reinforcement} & Meta-path based GE & Discovered meta-path over knowledge graph \\
			\hline
			KGAT~\cite{wang2019kgat} & GAT & Attention mechanism \\
			HAGERec~\cite{yang2020hagerec} & GAT & Attention mechanism \\
			\hline
		\end{tabular}
	\end{table}

	With the proliferation of GNN, researchers also make endeavors to improve the explainability of recommender systems with GNN's power of modeling logical relations. 
	He \textit{et al.}~\cite{he2015trirank} construct a heterogeneous graph with three kinds of nodes, including the user, the item, and the aspect~(the specific item property extracted from textual reviews). Therefore, they cast the recommendation task into a ternary relation ranking task, and propose TriRank with a high degree of explainability by explicitly modeling aspects in reviews. 
	
	Inspired by this work, the following research further explores rich information in dimensions of users and items, generally organized in the form of knowledge graph, in order to enhance the explainability~\cite{zhang20explainable}. 
	Ai \textit{et al.}~\cite{ai2018learning} construct a knowledge graph with entities, \textit{i.e.,} users and items, and relations, \textit{e.g.,} ``User A \textit{purchased} Item B \textit{belonging to} Category C''. Moreover, they embed each entity for recommendation and adopt the shortest relation path between a user and an item in the knowledge graph to indicate the recommendation explanations. 
	Different from separately utilizing embedding-based and path-based methods like Ai \textit{et al.}~\cite{ai2018learning}, Wang \textit{et al.}~\cite{wang2018ripplenet} proposed an end-to-end framework RippleNet which combines the two knowledge graph-aware recommendation methods together. Here, the knowledge graph contains the related knowledge of the recommended items, such as the type and author of a movie. In this way, explanations can be generated by the path between users' history to an item with high scores. 
	
	The meta-path-based utilization of knowledge graph can also benefit other specific recommendation tasks, e.g., sequential recommendation~\cite{huang2019explainable,wang2019explainable}. Huang \textit{et al.}~\cite{huang2019explainable} extract semantic meta-paths between a user and an item from knowledge graph to help sequential recommendation. Further, they encode and rank the meta-paths to generate the recommendation list, and these meta-paths also indicate the respective explanation. Similarly, Wang \textit{et al.}~\cite{wang2019explainable} also leverages knowledge graph to improve the performance of the sequential recommendation task and encode the meta-path between a user and an item with recurrent neural networks. Chen \textit{et al.}~\cite{chen2021temporal} further model the temporal meta-paths by capturing historical item features and path-defined context with neural networks.
	
	However, these meta-path-based solutions also face some challenges in terms of how to obtain these meta-paths~\cite{ma2019jointly, Xian2019reinforcement}. First,  since pre-defined meta-paths require extensive domain knowledge, Ma \textit{et al.}~\cite{ma2019jointly} jointly combine the discovery of inductive rules~(meta-paths) from the item-centric knowledge graph, which equips the framework with explainability and the learning of a rule-guided recommendation model. Moreover, to overcome the computational difficulties of enumerating all potential meta-paths, Xian \textit{et al.}~\cite{Xian2019reinforcement} replace the enumeration method with the reinforcement reasoning approach to identify proper meta-paths for scalability. 
	
	Besides meta-path-based solutions, Wang \textit{et al.}~\cite{wang2019kgat} propose a new method named Knowledge Graph Attention Network (KGAT), where the attention mechanism can offer explainability to some extent. Yang \textit{et al.}~\cite{yang2020hagerec} develop a hierarchical attention graph convolutional network to model higher-order relations in the heterogeneous knowledge graph, where the explainability is also dependent on the attention mechanism.
	\revise{Explainability in recommendation includes designing explainable model and finding explanations for those models without explainability (such as a black-box model). That is, the explainable model can naturally generate explainable recommendation results. As for GNN-based recommendation models for explainability, they can utilize paths on heterogenous graphs or knowledge graphs to generate recommendation. Since the model architecture is explicit and explainable, they are classified into the first category of explainability in recommendation.}
	\revise{The existing works of GNN-based recommendation for explainability mostly extract the relations between attributes or behaviors with recommendation results, which ignores the causality, which may lead to wrong explanations. Designing suitable causal inference methods into graph neural network-based models is important and promising.}
	
	\subsubsection{GNN for Fairness}
	\begin{figure}[t!]
		\centering 
		\includegraphics[width=0.75\textwidth]{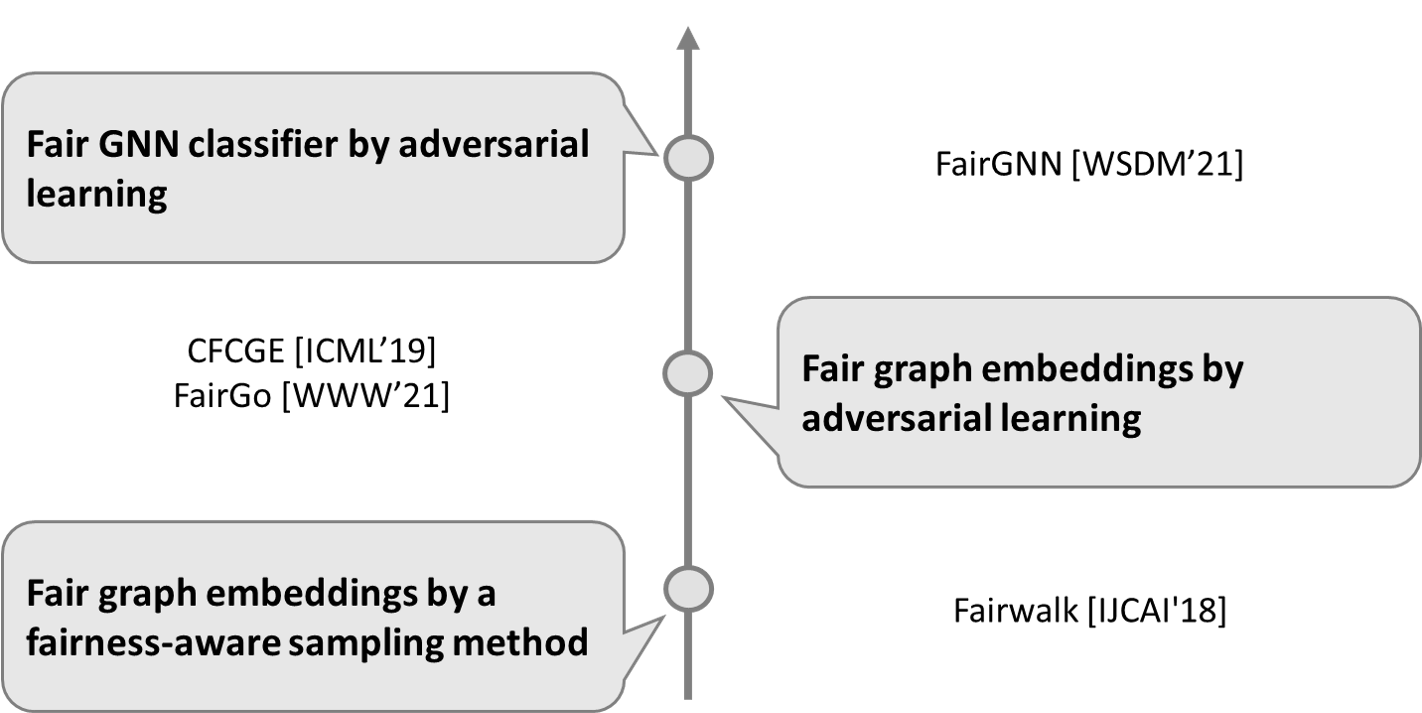} 
		\caption{Illustration of GNN models for fair recommendation.} \label{fig:fairness}
	\end{figure}
	\begin{table}[t!]
		\centering
		\small
		\caption{Details of GNN models for fair recommendation.}\label{tab:fair}
		\begin{tabular}{ccc}
			\hline
			Model & \revise{Network Design} & \revise{Achieving Fairness}\\
			\hline
			Fairwalk~\cite{Rahman2019fairwalk}  & Node2vec (GE) & learning fair graph embeddings by sampling \\
			CFCGE~\cite{bose2019compositional} & Invariant GE & learning fair graph embeddings by adversarial learning \\
			FairGo~\cite{wu2021learning} & HetGNN(GNN) &  learning fair graph embeddings by adversarial learning\\
			FairGNN~\cite{enyan2021say} & GraphSage(GNN) & learning fair GNN classifiers by adversarial learning\\
			\hline
		\end{tabular}
	\end{table}
	Despite the power of graph data in recommendation, it might inherit or even amplify discrimination and the societal bias in recommendation~\cite{enyan2021say,Rahman2019fairwalk,wu2021learning}. Past research has proven that compared with models that only adopt node attributes, the user unfairness is magnified due to the utilization of graph structures~\cite{enyan2021say}. 
	
	To curb the fairness issue in recommendation, some researchers propose to learn fair graph embeddings~\cite{Rahman2019fairwalk,bose2019compositional,wu2021learning}. Rahman \textit{et al.}~\cite{Rahman2019fairwalk} extends the well-known graph embedding method, node2vec~\cite{grover2016node2vec}, to a more fair version, Fairwalk, which can generate a more diverse network neighborhood representation for social recommendations by sampling the next node based on its sensitive attributes. Thus, all nodes' sensitive attributes are indispensable. Bose \textit{et al.}~\cite{bose2019compositional} proposes an adversarial framework to minimize the sensitive information in graph embeddings with a discriminator enforcing the fairness constraints. Moreover, considering the evaluations of fairness can vary, fairness constraints are flexible according to the task. However, similar to Fairwalk~\cite{Rahman2019fairwalk}, all nodes' sensitive attributes are required. Wu \textit{et al.}~\cite{wu2021learning} learn fair embeddings for recommendation from a graph-based perspective. They propose FairGo, which adopts a graph-based adversarial learning method to map embeddings from any recommendation models into a sensitive-information-filtered space, therefore eliminating potential leakage of sensitive information from both original recommendation embeddings and user-centric graph structures.
	Indeed, except for learning fair embeddings, Dai \textit{et al.}~\cite{enyan2021say} propose FairGNN, which learns fair GNN classifiers with limited known sensitive attributes in an adversarial learning paradigm with fairness constraints. Different from fair graph embeddings, fair GNN classifiers are to ensure node classification task~(rather than graph embeddings) independent of sensitive data. Moreover, they develop GNN-based sensitive data estimators to overcome the issue of missing sensitive data in the real world.
	
	\revise{In short, existing GNN-based recommendation models for achieving fairness always lead to performance drop to some extent, requiring solutions for addressing the trade-off dilemma.}
	
	\subsection{GNN for Specific Recommendation Applications}
	
	Graph neural networks are also widely used to handle the specific challenges in different applications of recommender systems.
	As for e-commerce/product recommendation, most of the existing works have been introduced in the above sections. 
	Li~\textit{et al.}~\cite{li2020hierarchical} propose to stack multiple GNN modules and use a deterministic clustering algorithm to help improve the efficiency of GNN in large-scale e-commerce applications.
	Liu~\textit{et al.}~\cite{liu2021item} propose to leverage the topology of item-relations for building graph neural networks in e-commerce recommendation.
	As for the Point-of-Interest recommendation, GGLR~\cite{chang2020learning} uses the sequence of POIs a user visited to construct the POI-POI graph, where the edges between POIs denote the frequency of users consecutively visiting two POIs. 
	Zhang~\textit{et al.}~\cite{zhang2021leveraging} propose to combine social networks and user-item interactions together, which deploys embedding aggregation on both social-connected users and visited POIs.
	As for news recommendation, Hu~\textit{et al.}~\cite{hu2020graph} propose to introduce the preference disentanglement~\cite{wang2022disentangled} into the user-news embedding propagation.
	
	There are many other papers for specific applications, but they can be well categorized into and covered by the corresponding stages, scenarios, and objectives, and thus we omit them here.
	\section{Open Problems and Future Directions}\label{sec::open-problems}

\subsection{\revise{More Complex GNNs}}

\subsubsection{\revise{Deeper GNN}} \label{deeper}
Due to the over-smoothing problem, more and more studies focus on properly increasing GNN's layers to capture higher-order connectivity correlations on graphs as well as improve models' performance~\cite{lai2020policy, zhou2020deepgnn,chen2020measuring,rong2019dropedge}. Despite these advancements, there is no universal solution for constructing very deep GNN like CNN, and relevant works propose different strategies. Lai \textit{et al.}~\cite{lai2020policy} develops a meta-policy to adaptively select the number of propagation for each graph node through training with reinforcement learning~(RL). The experiment results show that partial nodes require more than three propagation layers to boost model performance. Rong \textit{et al.}~\cite{rong2019dropedge} alleviates the over-smoothing problem by randomly removing graph edges, which acts as a message-passing reducer. Claudio \textit{et al.}~\cite{gallicchio2020fast} considers GNN as a dynamic system, and learned representations are the system's fixed points. Following this assumption, the transformation matrix in propagation is first fixed under stability conditions. Furthermore, only embeddings are updated in the learning procedure, leaving the matrix untrained. In this way, GNN can be trained faster and go deeper. Li \textit{et al.}~\cite{li2019deepgcns} transfer the concepts of residual/dense connections and dilated convolutions from CNNs to assist deeper GNNs.
As for future works, the performance leap compared with current shallow GNNs should be a fundamental problem in developing very deep GNNs, like groundbreaking works in the area of CNN~\cite{szegedy2015going,huang2017densely}. At the same time, the computation and time complexity must also be acceptable.

\subsubsection{\revise{Dynamic GNN}}

Existing GNN-based recommendation models are almost based on the static graph, as mentioned above, while there is plenty of \textit{dynamics} in recommender systems.
For example, in the sequential recommendation or session-based recommendation, the users' data is collected in a dynamic manner, naturally. In addition, modeling the dynamic user preferences is one of the most critical challenges in these recommendation scenarios.
In addition, the platform may dynamically involve new users, new products, new features, etc., which poses challenges to static graph neural networks.
Recently, dynamic graph neural networks~\cite{li2020dynamic,ma2020streaming} have attracted attention, which deploys embedding propagation operations on dynamically-constructed graphs.
Given the time-evolving property of recommender systems, the dynamic graph neural network-based recommendation model will be a promising research direction with broad applications in the real world.

\subsubsection{\revise{Efficiency and Scalability in Larger GNN}}
Early works on GNN follow the full-batch gradient descent algorithm, where the whole adjacency matrix is multiplied on the node embeddings during each inference step, which can not handle real-world recommender systems since the number of nodes and edges can reach a million-level scale.
A recent work~\cite{zhao2022joint} even proposes a GNN-based solution for both the searching and recommendation engines, which results in higher requirements of computational efficiency.
Hamilton \textit{et al.} propose GraphSAGE \cite{hamilton2017graphsage} which performs neighbor sampling and only updates the related sub-graph instead of the whole graph during each inference step.
The sampling strategy is also adopted in a few other works \cite{chiang2019cluster,chen2018fastgcn} which reduces the computation complexity of GNN and improves the scalability. 
Ying \textit{et al.} \cite{ying2018graph} successfully apply GraphSAGE to web-scale recommender systems, which can efficiently compute embeddings for billions of items.
Some other works~\cite{feng2022reinforcement} design specific query-recall acceleration strategies for an efficient recommendation.
In addition, a few open-source tools have been released which can accelerate the research and development of GNN-based recommendation, such as PyG \cite{fey2019fast}, DGL \cite{wang2019deep} and AliGraph \cite{zhu2019aligraph}.
Recently, a graph learning system with both effectiveness and scalability, PlatoGL~\cite{lin2022platogl}, has been proposed. It provides a practical solution to combine large-scale GNN with real-time recommendation, a fundamental problem for today's industrial recommendations.
We refer to another survey \cite{abadal2020computing} for details on computing and accelerating GNN.

Despite these existing approaches, achieving large-scale GNN-based recommendation is still a challenging task, especially in the ranking stage where thousands of features are involved, which results in a large and complicated heterogeneous graph.

\subsection{\revise{Advanced Machine Learning methods}}
\subsubsection{\revise{Self-supervised GNN}}
The direct supervision from interaction data is relatively sparse compared with the scale of the graph.
Therefore, it is necessary to include more supervision signals from the graph structure itself or the recommendation task.
For example, Yu \textit{et al.} \cite{yu2021MHCN_social} and Wu \textit{et al.} \cite{wu2021self} attempt to enhance GNN-based recommendation by designing auxiliary tasks from the graph structure with self-supervision.
Data augmentations such as node dropout are utilized to generate sample pairs for contrastive training.
We believe it is a promising future direction to leverage extra self-supervised tasks to learn meaningful and robust representations for GNN-based recommender systems.

\subsubsection{\revise{AutoML-enhanced GNN}}
Recommendation scenarios are diverse and vastly different from each other; thus, there exists no silver bullet GNN model that can generalize across all scenarios.
Recently, AutoML (Automated Machine Learning) \cite{yao2018taking} has been proposed, which can automatically design appropriate models for specific tasks.
For GNN-based recommendation, the search space is quite large, including multiple options for the neighbor sampler, aggregator, interaction function, etc.
Consequently, AutoML can, to a great extent, reduce human effort in discovering advanced model structures.
A few works \cite{gao2019graphnas,huan2021search} have been proposed which search to combine GNN layers and aggregate neighbors.
Therefore, designing AutoML algorithms to search for GNN-based recommender systems~\cite{wang2022profiling} is a promising future direction.

\subsection{\revise{Others}}

\subsubsection{\revise{About the Taxonomy}}
\revise{In this survey, we have presented a taxonomy for GNN-based recommendation, in which we discuss the existing works from four aspects: stage, scenario, objective, and application.
Here we discuss the important future directions following the presented taxonomy.
\begin{itemize}[leftmargin=*]
	\item \textbf{Stage.} The traditional recommendation divides the whole recommendation process into several stages mainly due to the too-large size of the original candidate pool (million or even billion). That is, different stages of recommender engines have different requirements for models. Currently, with the fast growth of computing power, the boundaries between different stages get more and more blurred, which can be understood in two folds. First, the model in the ranking stage starts to take a huge-size item pool and multiple user behaviors into consideration. Second, the multiple objectives in the re-ranking stage, such as diversity, can also be achieved by the model of the recall stage or ranking stage, making the re-ranking stage not so necessary.	\item \textbf{Scenario.} With the fast development of online information services, recommendation scenarios are also changing rapidly. For example, the user can interact with the recommender systems more actively by chatting with the robot, defined as a new scenario of conversational recommendation, which we will discuss in detail later. Recently, some works~\cite{zhao2022joint} consider the search and recommendation simultaneously, two core concepts of information retrieval, which try to design a unified graph neural network model to provide two kinds of services.
	\item \textbf{Objectives.} Existing works of GNN-based recommendation models have spent much effort in achieving various objectives, including accuracy, diversity, explainability, fairness, etc. However, they are still suffering from the trade-off dilemma, in which only one goal can be optimized simultaneously. Recently, some works of deep learning-based recommendation, such as MMOE~\cite{ma2018modeling}, PLE~\cite{tang2020progressive}, etc., studied how to deploy multiple separate models, each of which is designed towards one objective, and then optimize them together. We believe that these techniques in multi-objective learning can also be well used in addressing the trade-off in GNN-based recommendation models. 
	\item \textbf{Applications.} On the one hand, a powerful recommender system helps existing non-personalized applications attract more users and promote platform profits. On the other hand, relying on its benefit of both user-side and platform-side, the recommender system can motivate some new forms of applications, such as micro-video recommendation in TikTok. In these applications where content has abundant features, deploying cross-modal graph neural network models in both content understanding and user behavior modeling is a promising research direction. 
\end{itemize}
}

\subsubsection{\revise{KG-enhanced recommendation with GNN}}
Recent research has demonstrated the power of KG in recommendation by enriching the user-item bipartite graph with knowledge ~\cite{gao2020deep}. Specifically, the utilization of knowledge graph with GNN significantly addresses some practical issues in recommender systems~\cite{gao2020deep}, for example, cold start problem~\cite{fan2019metapath,zhang2019stargcn},
and dynamicity~\cite{sun2018recurrent}. Moreover, knowledge graph also offers a novel solution to some scenarios in recommendations, e.g., sequential recommendation~\cite{huang2019explainable,wang2019explainable}, and objectives, e.g., explainability~\cite{yang2020hagerec,wang2019kgat,park2022reinforcement}. 
There are already many works including KGCN~\cite{wang2019knowledge}, KGAT~\cite{wang2019kgat}, KGIN~\cite{wang2021learning}, etc. Currently, the majority of research first uses GNN to learn embeddings of knowledge graphs and then incorporate these embeddings into the recommendation model so that an end-to-end model can be trained~\cite{wang2020knowledge}. 
Therefore, we point out that the utilization of KG in recommendation with GNN can be further enhanced from the perspectives of data, scenario, and model. 
Specifically, the incorporated KG mainly records the rich item-item relations, e.g., Movie A belongs to Category B~\cite{gao2020deep}, but user-user knowledge is lacking in formal and plausible definition and thus substantially overlooked. Future work can consider creating user-centric KG based on abundant knowledge in sociology. Moreover, considering its success in scenarios, such as sequential recommendation~\cite{huang2019explainable,wang2019explainable}, it is promising that the leverage of KG could further enhance recommendation quality from the aspects that require more external knowledge, such as diversity and fairness. Further, existing methods of leveraging KG in recommendation cannot fully model complex relations between a user and a specific item or its attributes. Thus, designing a better framework to carve out these complex relations is another future direction.

\subsubsection{\revise{Conversational Recommendation with GNN}}
In existing recommender systems, there may exist the issue of information asymmetry that the system can only estimate users' preferences based on their \textit{historically collected} behavior data.
To address it, recently, conversational (interactive) recommendation researches~\cite{sun2018conversational, lei2020interactive,wu2022state,liu2022graph} proposed a new paradigm the user can interact with the system in conversations, and then new data can be dynamically collected.
Specifically, users can chat with the system to explicitly convey their consumption demands or offer positive/negative feedback on the recommended items.
As for future work, the advances in representation learning with graph neural networks can be combined with preference learning in the conversational recommendation.

	\section{Conclusion}\label{sec::conclusion}
	There is a rapid development of graph neural network models in the research field of recommender systems.
	This paper provides an extensive survey systematically presenting the challenges, methods, and future directions in this area.
	Not only the history of development and also the most recent advances are well covered and introduced.
	We hope this survey can well help both junior and experienced researchers in the relative areas.
	
	\begin{acks}
		This work is supported by the National Key Research and Development Program of China under grant 2020AAA0106000. 
		This work is also supported by the National Natural Science Foundation of China (U19A2079).
		This work is also supported by the fellowship of China Postdoctoral Science Foundation under grant 2021TQ0027 and 2022M710006.
	\end{acks}

	\bibliographystyle{ACM-Reference-Format}
	\bibliography{bibliography}

\end{document}